\documentclass[11pt,a4paper]{article}

\usepackage[T1]{fontenc}
\usepackage[utf8]{inputenc}
\usepackage[english]{babel}
\usepackage{csquotes}

\usepackage[
  a4paper,
  margin=1in,
  includeheadfoot,
  top=0.90in,
  headsep=20pt
]{geometry}

\usepackage{microtype}
\usepackage{parskip}
\usepackage{setspace}
\usepackage{ragged2e}
\AtBeginDocument{\justifying}

\usepackage{mathptmx}  

\usepackage{amsmath}
\usepackage{mathtools}

\usepackage[dvipsnames]{xcolor}
\definecolor{Accent}{HTML}{1D4ED8}
\definecolor{Accent2}{HTML}{0F766E}
\definecolor{Dark}{HTML}{111827}
\definecolor{SoftBG}{HTML}{F3F6FB}
\definecolor{Rule}{HTML}{CBD5E1}
\definecolor{Muted}{HTML}{475569}

\usepackage{hyperref}
\hypersetup{
  colorlinks=true,
  linkcolor=Accent,
  citecolor=Accent2,
  urlcolor=Accent
}
\usepackage[nameinlink,capitalise,noabbrev]{cleveref}

\usepackage{graphicx}
\usepackage{float}
\usepackage{caption}
\usepackage{booktabs}
\usepackage{multirow}
\usepackage{tabularx}
\usepackage{array}
\newcolumntype{Y}{>{\raggedright\arraybackslash}X}
\usepackage{siunitx}
\usepackage{pdflscape}
\usepackage{longtable}

\usepackage{enumitem}
\setlist[itemize]{topsep=4pt,itemsep=2pt,leftmargin=*}
\setlist[enumerate]{topsep=4pt,itemsep=2pt,leftmargin=*}

\usepackage[most]{tcolorbox}
\tcbset{
  enhanced,
  colback=SoftBG,
  colframe=Rule,
  boxrule=0.8pt,
  arc=2mm,
  left=10pt,right=10pt,top=8pt,bottom=8pt
}
\newtcolorbox{AbstractBox}{
  colframe=Rule,
  colback=SoftBG,
  title=\textbf{Abstract},
  fonttitle=\color{Dark},
  coltitle=Dark
}

\usepackage{titlesec}
\titleformat{\section}{\Large\bfseries\color{Dark}}{\thesection}{0.6em}{}
\titleformat{\subsection}{\large\bfseries\color{Dark}}{\thesubsection}{0.6em}{}
\titlespacing*{\section}{0pt}{0.6em}{0.35em}
\titlespacing*{\subsection}{0pt}{0.5em}{0.25em}

\usepackage{fancyhdr}
\renewcommand{\headrulewidth}{0.5pt}
\renewcommand{\headrule}{\hbox to\headwidth{\color{Rule}\leaders\hrule height \headrulewidth\hfill}}
\fancypagestyle{lscapestyle}{%
  \fancyhf{}%
  \renewcommand{\headrulewidth}{0pt}%
}

\usepackage[numbers,sort&compress]{natbib}

\newcommand{\orcidicon}{\textsuperscript{iD}}

\newcommand{\authorname}[4]{%
  \textbf{\color{Dark}#1}\textsuperscript{#3}%
  \href{https://orcid.org/#2}{\textcolor{Accent}{\,\orcidicon}}%
  \ifnum#4=1\relax\textsuperscript{*}\fi
}

\newcommand{\corremail}[1]{%
  {\small\color{Muted}\textsuperscript{*} Corresponding author: \href{mailto:#1}{#1}\par}
}

\newcommand{\papertitle}{Maritime Connectivity Vulnerability Index: Construction, Patterns, and Validation Across 185 Economies, 2006--2025}
\newcommand{\paperkeywords}{maritime vulnerability; liner shipping; composite indicator; connectivity divide; port concentration; Monte Carlo uncertainty analysis; supply-side disruption; SIDS; LDC; panel data; UNCTAD; LSCI; LSBCI; Herfindahl--Hirschman Index}

\makeatletter
\renewcommand{\maketitle}{%
  \begingroup
  \thispagestyle{empty}
  \vspace*{-1.20cm}
  {\color{Accent}\rule{\linewidth}{1.4pt}}\par
  \vspace{0.15cm}
  {\bfseries\LARGE\color{Dark}\papertitle\par}
  \vspace{0.22cm}
  {\color{Rule}\rule{\linewidth}{0.5pt}}\par
  \vspace{0.22cm}
  \begin{center}
    {\normalsize
      \authorname{Mohamed Bouka}{0009-0005-8502-9885}{1}{1},
      \authorname{Moulaye Abdel Kader Moulaye Ismail}{0009-0005-6738-9975}{1}{0}
      \par
    }
    \vspace{0.12cm}
    {\small\color{Muted}
      \textsuperscript{1}Research Unit of Governance of Institutions, Faculty of Economics and Management, University of Nouakchott, Nouakchott, 5026, Mauritania \\
      \par
    }
    \vspace{0.08cm}
    \corremail{mohamedbouka50@gmail.com}
    {\small\color{Muted}\href{mailto:Moulaye.abdelkader@gmail.com}{Moulaye.abdelkader@gmail.com}\par}
  \end{center}
  \vspace{0.15cm}
  \endgroup
  \pagestyle{fancy}
}
\makeatother

\begin{document}
\maketitle

\begin{AbstractBox}
Recent disruptions at major maritime chokepoints have exposed the structural fragility of liner shipping networks, particularly for countries with limited service portfolios. While existing indicators such as the Liner Shipping Connectivity Index (LSCI) measure how well countries are connected, no composite framework quantifies how structurally vulnerable that connectivity is from a supply-side perspective. This paper proposes the Maritime Connectivity Vulnerability Index (MCVI), capturing three dimensions of supply-side fragility mapped to distinct UNCTAD data sources: (i)~low overall connectivity (LSCI), (ii)~weak bilateral integration (LSBCI), and (iii)~port infrastructure concentration (PLSCI). The index covers 185~economies over 2006--2025 using pooled fractional rank normalization and equal-weight aggregation, constructed exclusively from publicly available data. Results reveal substantial disparities: SIDS exhibit a mean vulnerability 0.234~points above non-SIDS economies, a gap that has widened over two decades from 0.232 to 0.249. A modest global decline of 4.2\% is observed, but improvements are unevenly distributed. Dimension decomposition identifies port concentration as the dominant vulnerability channel for nearly 40\% of economies (72 of 185, 39\%), establishing infrastructure diversification as a distinct policy priority. Rankings are highly stable across alternative weighting schemes (Spearman $\rho > 0.95$), normalization methods ($\rho = 0.97$--$0.999$), and PCA-derived weights ($\rho = 0.999$), with Monte Carlo simulation (1{,}000~replications) confirming $\rho > 0.95$ in every realization. External validation shows strong negative correlation with the World Bank Logistics Performance Index ($\rho \approx -0.61$ across seven waves) and positive correlation with ad valorem maritime freight rates ($\rho \approx +0.32$). Panel regressions confirm a negative association with GDP per capita and identify a vulnerability paradox of trade openness among small, trade-dependent economies, with a Hausman test strongly favoring fixed effects ($H = 370.76$, $p < 0.001$). Predictive validity tests show that pre-crisis MCVI scores predict trade losses during supply-side shocks ($\rho = -0.25$, $p < 0.005$ for COVID-19), while the contrasting pattern during the 2008--2009 demand shock ($\rho = +0.23$, $p = 0.01$) validates the supply-side specificity of the index.
\end{AbstractBox}

\vspace{0.15cm}

\newpage

\section{Introduction}\label{sec:introduction}

Maritime transport constitutes the primary physical infrastructure of contemporary international trade, with containerized liner shipping forming its dominant logistics architecture for manufactured goods, intermediate inputs, and consumer products. The distribution of liner services across countries is structurally uneven, reflecting differences in cargo volumes, port infrastructure, geographic positioning, and integration into alliance-driven service networks \citep{guo2024, tsantis2026}. For countries with limited trade volumes or constrained port capacity, this unevenness translates into narrow service portfolios, dependence on transshipment hubs, and reduced routing flexibility, which are structural characteristics that condition exposure to service disruptions \citep{unctad2024rmt, xu2024}.

The practical significance of such structural dependencies has been underscored by a series of concurrent disruptions affecting major maritime chokepoints in recent years. The closure of the Strait of Hormuz following the February--March 2026 Iran conflict effectively suspended transit through one of the world's most critical energy and trade corridors, with direct consequences for container shipping operations in the Persian Gulf region \citep{hormuz2026}. Since late 2023, security threats in the Red Sea have led major carriers to reroute Asia--Europe and Asia--West Africa services via the Cape of Good Hope, increasing transit times and freight costs along corridors used by low-volume trading economies \citep{unctad2024rmt}. Restrictions on Panama Canal transits due to water-level shortages in 2023--2024 further compounded the routing constraints for the Atlantic and Pacific trade lanes \citep{panama2024}. The simultaneous disruption of three major chokepoints represents an unprecedented configuration in the modern history of containerized shipping and has intensified attention to the structural fragility of national maritime connectivity.

Within this context, several institutional and academic indicators have been developed to characterize maritime network configurations. The Liner Shipping Connectivity Index (LSCI), produced quarterly by UNCTAD and MDS Transmodal, measures the extent to which countries are integrated into global liner shipping networks based on six service-related components \citep{hoffmann2005, unctad2024lsci}. The Liner Shipping Bilateral Connectivity Index (LSBCI) extends this framework to country pairs, providing an aggregate score of bilateral maritime connection quality \citep{fugazza2013}. The Container Port Performance Index (CPPI), developed by the World Bank and S\&P Global Market Intelligence, assesses operational efficiency at the port level based on vessel time in port \citep{worldbank2024cppi}. More recently, \citet{arvis2026} proposed the Global Supply Chain Stress Index -- Maritime (GSCSI-M), a metric of realized disruption measured through delayed container capacity at ports.

These indicators provide valuable diagnostic information on connectivity levels, bilateral integration, port performance and observed stress. However, they do not directly quantify the structural vulnerability of a country's liner shipping connectivity -- that is, its susceptibility to experiencing severe logistical consequences from a disruption, given the configuration of its maritime service supply. Connectivity and vulnerability are conceptually distinct. A country may exhibit moderate connectivity while maintaining diversified bilateral connections and multiple port gateways (low vulnerability) or, conversely, may achieve reasonable connectivity scores while depending on few bilateral partners and one dominant port (high vulnerability). Existing indicators capture aspects of the configuration but do not aggregate them into a unified vulnerability measure.

\citet{guo2026} constructed a country-level vulnerability index for maritime bulk commodity supply chains using a weighted optimization model that integrates supplier-country and maritime route risks. Their framework addresses vulnerability in the context of strategic bulk commodities (iron ore and coal) and relies on commercial data sources. No analogous composite index currently exists for containerized liner shipping connectivity based on the publicly available data.

This study proposes the Maritime Connectivity Vulnerability Index (MCVI), a composite indicator designed to measure the structural vulnerability of countries' liner shipping connectivity from a supply-side perspective. The index captures three conceptually distinct dimensions of structural fragility, each mapped to a unique UNCTAD data source: (i)~low overall connectivity (LSCI), (ii)~weak bilateral integration (LSBCI), and (iii)~port infrastructure concentration (PLSCI). The architecture of one dimension per source dataset ensures conceptual clarity, minimizes redundancy, and parallels the design of established composite indices such as the Human Development Index \citep{undp2024hdr}. The MCVI is constructed exclusively from publicly available UNCTAD datasets, ensuring full transparency and replicability.

The main objective of this study is to develop and apply a composite indicator of supply-side vulnerability in liner shipping connectivity across countries. This objective was operationalized through the following research questions:

\begin{description}
\item[RQ1 -- Index Construction:] How can publicly available UNCTAD maritime indicators be combined into a composite index that captures structural vulnerability in liner shipping connectivity?
\item[RQ2 -- Cross-Country Vulnerability Patterns:] What is the global distribution of maritime connectivity vulnerability, and which country groups (SIDS, LDCs, and small coastal economies) exhibit the highest structural fragility?
\item[RQ3 -- Temporal Dynamics:] Has the structural vulnerability of liner shipping connectivity increased or decreased over the period 2006--2025, and do temporal patterns differ across country income groups or geographic regions?
\item[RQ4 -- Decomposition and Vulnerability Profiles:] For countries exhibiting high MCVI scores, which dimensions of structural vulnerability contribute most to the overall index, and do dominant vulnerability profiles differ systematically across country groups?
\item[RQ5 -- Sensitivity and Robustness:] To what extent is the country ranking produced by the MCVI sensitive to alternative weighting schemes and aggregation methods?
\end{description}

This study contributes to the emerging literature on maritime resilience and connectivity assessment by offering a standardized, data-driven diagnostic tool that complements existing connectivity and performance metrics from an explicit vulnerability perspective. The MCVI framework is designed to be applicable across countries and time periods, enabling comparative and longitudinal analyses of structural fragility in containerized maritime systems.

The remainder of this paper is organized as follows. \Cref{sec:related} reviews the related work. \Cref{sec:framework} presents the conceptual framework. \Cref{sec:data} describes the data and methods. \Cref{sec:results} reports the results. \Cref{sec:discussion} discusses the findings. \Cref{sec:conclusion} concludes.

\section{Related Work}\label{sec:related}

The literature relevant to this study spans four interrelated strands: (i) maritime connectivity measurement, (ii) vulnerability and resilience in maritime networks, (iii) maritime connectivity of vulnerable economies, and (iv) composite indicator methodology.

\subsection{Maritime Connectivity Measurement}\label{sec:connectivity}

The measurement of maritime connectivity has evolved from descriptive port-system assessments toward formal indicator-based frameworks. The LSCI, introduced by UNCTAD in 2004 and jointly produced with MDS Transmodal, measures a country's integration into global liner shipping networks based on six components: the number of ships, total deployed container-carrying capacity, maximum vessel size, number of scheduled services, number of liner companies, and number of countries connected through direct services \citep{hoffmann2005, unctad2024lsci}. The index is published quarterly for 187 economies and has become a standard reference in trade facilitation research, with demonstrated associations between connectivity levels and bilateral trade costs \citep{fugazza2017, hoffmann2020}.

The LSCI was revised in March 2024 to adjust for structural changes in vessel deployment patterns, particularly the increased role of ship size relative to other components \citep{unctad2024lscirevision}. Both the original and revised formulations aggregate six components into a single score, which captures overall connectivity but does not distinguish between qualitatively different configurations that may underlie similar scores.

At the bilateral level, the LSBCI measures the quality of maritime connections between country pairs based on five underlying components: the number of transshipments required, the number of common direct connections, the number of common connections with one transshipment, the level of service competition, and the size of the largest vessel on the weakest connecting route \citep{fugazza2013, fugazza2019}. The LSBCI is published as an aggregate bilateral score for each country pair. \citet{fugazza2017} demonstrated that the absence of a direct maritime connection is associated with export reductions between 42\% and 55\%, and that each additional transshipment corresponds to approximately a 40\% decline in bilateral export value.

At the port level, the PLSCI extends the LSCI framework to individual ports, covering more than 1,300 ports worldwide \citep{unctad2024lsci}. \citet{martinezmoya2020} developed a Foreland Port Connectivity Index for Spanish ports, disaggregating connectivity by destination market, while \citet{martinezmoya2024} applied benefit-of-the-doubt methods to assess connectivity and competitiveness in Mediterranean container ports.

Beyond UNCTAD indicators, several studies have proposed alternative connectivity measures. \citet{bartholdi2016} introduced a container port connectivity index based on network topology. \citet{wang2022cci} developed a Composite Connectivity Index (CCI) combining port centrality measures with connectivity scores using two-stage data envelopment analysis. \citet{mishra2021} revisited the LSCI using stochastic multicriteria acceptability analysis (SMAA), producing probabilistic rankings that account for weight variability. These contributions refine the measurement of connectivity but do not extend the framework toward vulnerability assessment.

\subsection{Vulnerability and Resilience in Maritime Networks}\label{sec:vulnerability}

At the network level, \citet{ducruet2022entropy} proposed an entropy-based model to quantify the vulnerability of the Asia--Europe maritime transportation network. \citet{calatayud2017} modeled vulnerability as a function of multiplex network configuration, simulating targeted attacks on strategic nodes. \citet{wang2023glsn} developed an integrated framework for assessing the efficiency and vulnerability of the global liner shipping network, finding that vulnerability increases nonlinearly under cascading failures of critical nodes. \citet{liu2024cascading} examined cascading failure vulnerability in the Maritime Silk Road container shipping network.

At the port level, \citet{verschuur2023multihazard} performed a global asset-level risk analysis covering 1,350 ports, finding that 86\% are exposed to more than three hazards and that trade risk is particularly high in SIDS. \citet{verschuur2021criticality} quantified port criticality for global supply chains, finding that low-income countries and small islands are 1.5 and 2.0 times more reliant on their ports compared to the global average. \citet{yang2021pva} developed a port vulnerability assessment framework combining fuzzy theory and evidential reasoning.

Empirical disruption studies include \citet{notteboom2021}, who compared the 2008--2009 financial crisis and COVID-19 impacts on container shipping. The Red Sea crisis has generated growing evidence on network reconfiguration under geopolitical disruption \citep{redsea2024mel}. \citet{arvis2026} proposed the GSCSI-M measuring realized disruption at ports. \citet{guo2026} constructed a country-level vulnerability index for maritime bulk commodity supply chains, representing the closest antecedent to the present study but differing in scope (bulk commodities) and data requirements (commercial sources). More recently, \citet{fan2025vtp} proposed the Vessel--Trade route--Port-call (VTP) framework for supply-side analysis of shipping network disruption amidst geopolitical tensions, using shipping services as the unit of analysis to study alliance behavior during the Red Sea crisis. The VTP framework and the MCVI are conceptually complementary: the former operates at the service-level with vessel-tracking granularity to analyze carrier responses to realized disruptions, whereas the MCVI operates at the country-level with publicly available scheduled-service data to quantify structural \emph{ex ante} vulnerability.

No existing study aggregates multiple dimensions of structural vulnerability into a composite country-level index for containerized liner shipping based on publicly available indicators.

\subsection{Maritime Connectivity of Vulnerable Economies}\label{sec:sids}

UNCTAD \citep{unctad2024rmt} reported that SIDS are on average ten times less connected to global shipping networks compared to non-SIDS countries. \citet{verschuur2021criticality} found that 40 ports add more than 10\% of domestic output of the economies they serve, predominantly in small island states. \citet{fugazza2017} demonstrated that connectivity gaps translate directly into trade cost penalties. \citet{rojon2021carbon} showed that the relationship between transport costs and trade performance is negative and elastic, with SIDS and LDCs being disproportionately affected. \citet{mohan2023} found that hurricanes reduce goods exports by 20\% in Eastern Caribbean SIDS.

At the regional level, UNCTAD \citep{unctad2022asean} documented persistent connectivity gaps between hub economies and peripheral island states. \citet{guerrero2022africa} visualized changes in maritime connectivity of African countries, and \citet{mohamedcherif2016} examined regional integration across the Maghreb seaport system.

\subsection{Composite Indicator Methodology}\label{sec:cimethodology}

The OECD/JRC Handbook \citep{oecd2008handbook} provides the foundational reference framework for composite indicator construction. \citet{greco2018} examined the methodological landscape of composite indices, identifying equal weighting as the most commonly applied approach while noting that PCA and DEA offer alternatives. Min-max normalization is widely used \citep{oecd2008handbook, greco2018}, though it is sensitive to outliers \citep{sensitivity2024minmax}. Additive aggregation assumes full compensability; geometric aggregation reduces this \citep{geometric2021}. Monte Carlo simulation, as in the Proteus food security index \citep{proteus2020}, offers comprehensive uncertainty propagation.

In the maritime domain, \citet{karakitsos2021} demonstrated the feasibility of a composite maritime-sector index. The LSCI itself is a composite indicator with equal weights, and the SMAA revisions by \citet{mishra2021} address weight uncertainty. The present study extends this line by constructing a composite indicator targeting vulnerability rather than connectivity.

In summary, while the measurement of maritime connectivity has advanced substantially and network vulnerability has received growing attention, no study has aggregated multiple dimensions of structural supply-side vulnerability into a composite country-level index for containerized liner shipping constructed from publicly available data with consistent temporal coverage. The present study addresses this gap.

\section{Conceptual Framework}\label{sec:framework}

\subsection{Defining Supply-Side Vulnerability}\label{sec:definition}

The structural vulnerability of a country's liner shipping connectivity is defined as the susceptibility of that country to experiencing severe logistical consequences from a disruption to its maritime service supply, given the configuration of its liner shipping network. This definition is scoped to the supply side---the characteristics of available liner services---rather than demand-side economic characteristics.

Three conceptual distinctions are central. First, vulnerability is distinct from connectivity: the LSCI captures integration, while the MCVI captures the fragility of that integration. Second, vulnerability is distinct from realized disruption. The MCVI is an \emph{ex ante} structural property of the shipping network configuration (scheduled services, bilateral partners, port concentration), measurable at any point in time regardless of whether a disruption has occurred. In contrast, indicators such as the GSCSI-M \citep{arvis2026} are \emph{ex post} metrics derived from AIS vessel-tracking data that quantify stress only after disruptions materialize. The two perspectives are complementary: a country with high \emph{ex ante} MCVI (structurally fragile service portfolio) is hypothesized to experience larger \emph{ex post} GSCSI-M stress when a supply shock occurs, and \Cref{sec:events} provides empirical support for this prediction using COVID-19 as a natural experiment. Third, the MCVI does not capture demand-side factors (GDP dependence on maritime trade, essential import composition) or chokepoint-specific exposure, which would require vessel-tracking data beyond publicly available sources.

\subsection{Dimensions of Supply-Side Vulnerability}\label{sec:dimensions}

The MCVI is structured around three dimensions, each operationalized using a distinct UNCTAD dataset. Throughout, $i$ denotes a country and $t$ a year (first-quarter observation). This one-dimension-per-source architecture parallels the design of established composite indices \citep{undp2024hdr, oecd2008handbook} and ensures that each dimension captures a conceptually distinct facet of vulnerability.

\subsubsection{Dimension 1: Low Overall Connectivity (LSCI)}\label{sec:dim1}

Countries with low LSCI scores have fewer ships, less capacity, fewer services, and fewer direct connections \citep{hoffmann2005, hoffmann2020}. Let $L_{i}^{t}$ denote the LSCI score of country~$i$ in year~$t$. Lower LSCI implies higher vulnerability.

\subsubsection{Dimension 2: Weak Bilateral Integration (LSBCI)}\label{sec:dim2}

A country's bilateral connectivity reflects both the quality and the diversity of its maritime partnerships. D2 is a composite of two indicators derived from the LSBCI dataset. The first is the mean bilateral connectivity quality:
\begin{equation}\label{eq:meanlsbci}
\bar{B}_{i}^{t} = \frac{1}{n_{i}^{t}} \sum_{j=1}^{n_{i}^{t}} B_{ij}^{t}
\end{equation}
where $B_{ij}^{t}$ is the LSBCI score for the pair $(i,j)$ and $n_{i}^{t}$ is the number of bilateral partners. The second is the count of bilateral partners~$n_{i}^{t}$ itself, measuring network diversity. These two sub-indicators are normalized separately and averaged to form D2, capturing both the depth and breadth of bilateral integration. Lower values on either sub-indicator imply higher vulnerability.

The decision to combine quality and diversity into a single dimension, rather than treating them as separate dimensions, is motivated by two considerations. First, the two sub-indicators are empirically near-redundant (Spearman $\rho \approx 0.99$ after normalization), meaning that countries with high-quality bilateral connections almost invariably maintain numerous partnerships. Treating them as separate dimensions would effectively double-count the same underlying factor. Second, both derive from the same LSBCI dataset, and the one-dimension-per-source principle avoids artificial inflation of the number of dimensions \citep{oecd2008handbook}.

\subsubsection{Dimension 3: Port Infrastructure Concentration (PLSCI)}\label{sec:dim3}

A single-port country is structurally more fragile because disruption at its dominant port has no substitute \citep{verschuur2021criticality}. D3 is computed as the Herfindahl--Hirschman Index (HHI) over port-level PLSCI scores within each country:
\begin{equation}\label{eq:hhi}
\mathrm{HHI}_{i}^{t} = \sum_{k=1}^{K_i} \left(\frac{p_{i,k}^{t}}{\sum_{j=1}^{K_i} p_{i,j}^{t}}\right)^{2}
\end{equation}
where $p_{i,k}^{t}$ denotes the PLSCI score of port~$k$ in country~$i$. Single-port countries receive $\mathrm{HHI}=1$ (maximum concentration). Approximately 40\% of observations exhibit $\mathrm{HHI}=1.0$, reflecting the empirical reality that many economies depend on a single port facility---a structural characteristic rather than a measurement artifact \citep{verschuur2023multihazard}.

\subsection{Normalization}\label{sec:normalization}

Indicators are normalized using pooled fractional ranks across the entire panel of $N$~country-year observations. For each raw indicator, the normalized value is:
\begin{equation}\label{eq:rank}
V_{d,i}^{t} = \frac{R_{d,i}^{t}}{N}
\end{equation}
where $R_{d,i}^{t}$ is the rank of observation $(i,t)$ among all $N=3{,}476$ observations (using average ranks for ties), with the sign adjusted so that higher values indicate greater vulnerability.

Pooled normalization---ranking each country-year against the entire panel rather than within each year---ensures both cross-sectional comparability and temporal consistency: if global connectivity improves over time, the per-year mean MCVI declines accordingly. This approach offers three advantages over within-year min-max normalization: (i)~it eliminates boundary effects, as no observation attains exactly 0 or~1; (ii)~it is robust to outliers without requiring ad hoc winsorization; and (iii)~it is consistent with recommendations in the OECD Handbook for ordinal normalization in panel settings \citep{oecd2008handbook, greco2018}.

\subsection{Index Aggregation}\label{sec:aggregation}

The MCVI is computed as the arithmetic mean of the three normalized dimensions:
\begin{equation}\label{eq:mcvi}
\mathrm{MCVI}_{i}^{t} = \frac{1}{3}\bigl(D_{1,i}^{t} + D_{2,i}^{t} + D_{3,i}^{t}\bigr)
\end{equation}
where $D_{2,i}^{t} = \tfrac{1}{2}(D_{2a,i}^{t} + D_{2b,i}^{t})$ is the average of the two LSBCI-derived sub-indicators. Equal weights are adopted on the grounds that each dimension captures a conceptually distinct facet of vulnerability measured from an independent data source \citep{oecd2008handbook}. The MCVI takes values in approximately $[0,1]$, where higher values indicate higher structural vulnerability. The sensitivity of results to alternative weighting is assessed in \Cref{sec:robustness}.

\section{Data and Methods}\label{sec:data}

\subsection{Data Sources}\label{sec:sources}

Three publicly available UNCTAD datasets are used, accessed through the UNCTADstat Data Centre \citep{unctad2024lsci}. \Cref{tab:data} summarizes the data sources.

\begin{table}[H]
\centering
\caption{UNCTAD datasets used in MCVI construction.}\label{tab:data}
\begin{tabular}{l l l l p{4.5cm}}
\toprule
Dimension & Source & Level & Period & Indicator \\
\midrule
D1: Connectivity & LSCI & Country & 2006--2025 & Country LSCI score (inverted) \\
D2: Bilateral integration & LSBCI & Country pair & 2006--2025 & Mean LSBCI + partner count (composite, inverted) \\
D3: Port concentration & PLSCI & Port & 2006--2025 & HHI of port PLSCI scores (direct) \\
\bottomrule
\end{tabular}
\end{table}

The LSBCI is harmonized at Q1 frequency; all dimensions are therefore observed annually using Q1 values. The resulting panel comprises 3,476 country-year observations across 185~economies and 20~years (2006--2025). The panel is mildly unbalanced, with country counts ranging from 168 (2007) to 177 (2019--2020). No imputation is applied.

\subsection{Variable Construction}\label{sec:variables}

\textbf{D1.} The LSCI score is used directly; higher values indicate lower vulnerability. \textbf{D2.} Two sub-indicators are derived from the LSBCI: the mean bilateral index value across all partners (\Cref{eq:meanlsbci}) and the number of bilateral connections. Each is normalized separately via pooled fractional ranks and then averaged. \textbf{D3.} The HHI is computed over PLSCI scores of all ports within country~$i$ (\Cref{eq:hhi}); single-port countries receive $\mathrm{HHI}=1$.

\subsection{Analytical Procedures}\label{sec:procedures}

The analysis addresses each research question through dedicated statistical procedures.

\textbf{RQ1 (Index Properties).} Inter-dimension pairwise correlations are computed on the full panel. Principal component analysis (PCA) is applied to the standardized dimension scores to assess the latent structure and effective dimensionality of the index \citep{greco2018}.

\textbf{RQ2 (Cross-Country Patterns).} Time-averaged MCVI scores are compared across four country classifications: SIDS (as classified by UNCTAD), LDCs, LLDCs, and five geographic regions based on the UN~M49 standard (Africa, Americas, Asia, Europe, Oceania). Descriptive statistics (mean, standard deviation) are reported for each group.

\textbf{RQ3 (Temporal Dynamics).} The global MCVI trend is estimated by ordinary least squares regression of the annual cross-country mean on a linear time index. Segment-level trends are reported separately for SIDS vs.\ non-SIDS economies. Rank-order stability is assessed using Spearman rank correlations between consecutive years. Country-level volatility is measured as the standard deviation of each country's MCVI over the sample period.

\textbf{RQ4 (Decomposition and Clustering).} For each country, the dominant dimension is identified as $\arg\max_{d}\bar{D}_{d,i}$ over the time-averaged scores. $K$-means clustering is applied to the standardized time-averaged dimension profiles $(\bar{D}_1, \bar{D}_2, \bar{D}_3)$ for $k \in \{2,\ldots,6\}$, with the optimal number of clusters selected by silhouette analysis \citep{oecd2008handbook}.

\textbf{RQ5 (Robustness and External Validation).} Internal robustness is assessed through four tests: (i)~PCA-derived weights using the absolute loadings of the first principal component, normalized to sum to unity; (ii)~leave-one-dimension-out analysis comparing two-dimension MCVI variants with the baseline via Spearman~$\rho$; (iii)~within-year rank normalization as alternative to pooled ranks; and (iv)~pooled min-max normalization without winsorization. External validation uses pooled OLS regressions with standard errors clustered at the country level, regressing MCVI on log GDP per capita (World Bank WDI), log trade openness (trade-to-GDP ratio), and SIDS and LDC dummies. Fixed-effects and random-effects specifications are estimated as sensitivity checks, with the Hausman specification test used to adjudicate between the two \citep{oecd2008handbook}.

\textbf{Monte Carlo Uncertainty Analysis.} Following Step~7 of the OECD/JRC Handbook \citep{oecd2008handbook} and the framework of \citet{saisana2005}, a Monte Carlo simulation propagates three sources of methodological uncertainty simultaneously: (i)~aggregation weights, sampled from a symmetric Dirichlet distribution ($\alpha = 20$) centered on equal weights, producing moderate perturbations around $w = (1/3, 1/3, 1/3)$; (ii)~measurement noise, applied as a uniform $\pm 5\%$ multiplicative perturbation to all raw indicator values; and (iii)~normalization choice, randomly switching between pooled panel ranks (baseline) and within-year ranks with probability 0.30. For each of 1{,}000~simulations, the MCVI is recomputed for all 3{,}476~observations and countries are re-ranked. Rank stability is assessed via Spearman correlations with the baseline ranking, 95\% confidence intervals on country ranks, and a Sobol-like variance decomposition attributing total rank uncertainty to each source \citep{saltelli2008}.

\textbf{Predictive Validity (Disruption Event Analysis).} To assess whether the MCVI has predictive power for realized disruption impacts, two natural experiments are exploited: the COVID-19 pandemic (2019--2020), which constituted a global supply-side shock to maritime logistics \citep{notteboom2021, guerrero2022covid}, and the 2008--2009 global financial crisis, which represented a demand-side shock \citep{notteboom2021}. For each event, countries are ranked by their pre-crisis MCVI and divided into quartiles. The change in trade openness (trade-to-GDP ratio, World Bank WDI) between the pre-crisis and crisis year is computed for each country. Spearman rank correlations between pre-crisis MCVI and trade change are reported, and Mann-Whitney $U$-tests compare the most and least vulnerable quartiles. The Red Sea crisis (2023--2024) is examined as a supplementary case to assess route-specific disruption effects not captured by the MCVI.

All computations are performed in Python~3.12 using \texttt{pandas}, \texttt{numpy}, \texttt{scipy}, \texttt{scikit-learn}, and \texttt{linearmodels}.

\section{Results}\label{sec:results}

\subsection{Index Properties (RQ1)}\label{sec:rq1}

The MCVI ranges from 0.007 (China, 2021) to 0.932 (Norfolk Island, 2021), with a panel mean of 0.500 and standard deviation of 0.255. No observation attains the theoretical bounds of 0 or~1, confirming that pooled rank normalization avoids boundary saturation. \Cref{tab:descstats} reports descriptive statistics for each dimension.

\begin{table}[H]
\centering
\caption{Descriptive statistics of MCVI dimensions ($N=3{,}476$).}\label{tab:descstats}
\begin{tabular}{l S[table-format=1.4] S[table-format=1.4] S[table-format=1.4] S[table-format=1.4]}
\toprule
& {Min} & {Max} & {Mean} & {Std.\ dev.} \\
\midrule
D1 (LSCI) & 0.0000 & 0.9991 & 0.4999 & 0.2887 \\
D2 (LSBCI composite) & 0.0006 & 0.9944 & 0.4999 & 0.2876 \\
D3 (HHI ports) & 0.0003 & 0.8048 & 0.5001 & 0.2800 \\
MCVI & 0.0071 & 0.9318 & 0.5000 & 0.2553 \\
\bottomrule
\end{tabular}
\end{table}

\begin{figure}[H]
\centering
\includegraphics[width=0.55\textwidth]{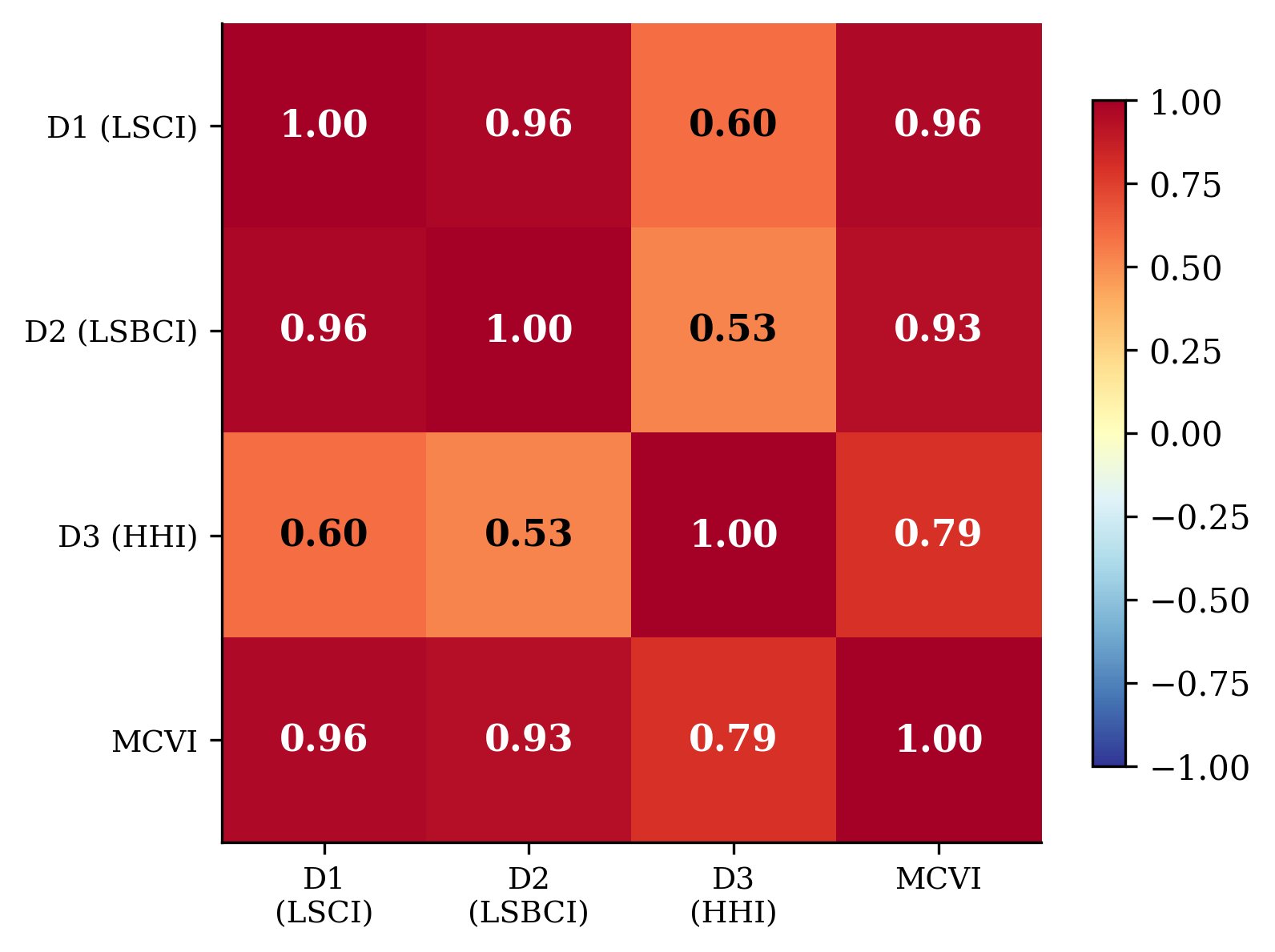}
\caption{Pairwise correlation matrix of MCVI dimensions.}\label{fig:correlation}
\end{figure}

The three dimensions exhibit pairwise correlations of $\rho(\text{D1},\text{D2}) = 0.964$, $\rho(\text{D1},\text{D3}) = 0.599$, and $\rho(\text{D2},\text{D3}) = 0.529$ (\Cref{fig:correlation}). The high D1--D2 correlation reflects the empirical regularity that countries with strong national connectivity also maintain diverse bilateral linkages---a substantive finding rather than a methodological artifact. Dimension~D3 provides a distinct axis of variation.

PCA on the full panel ($N=3{,}476$ country-year observations) yields explained variance shares of $\approx$80.5\%, $\approx$18.5\%, and $\approx$1.1\%. The first principal component loads approximately equally on all three dimensions (0.62, 0.61, 0.49), representing general connectivity--vulnerability. The second component opposes D3 (loading 0.87) to D1 and D2 ($-0.29$ and $-0.41$), capturing port concentration as a distinct vulnerability axis. This two-factor structure confirms that the MCVI captures two interpretable latent constructs: overall maritime connectivity and infrastructure concentration.

\subsection{Cross-Country Patterns (RQ2)}\label{sec:rq2}

\begin{figure}[H]
\centering
\includegraphics[width=\textwidth]{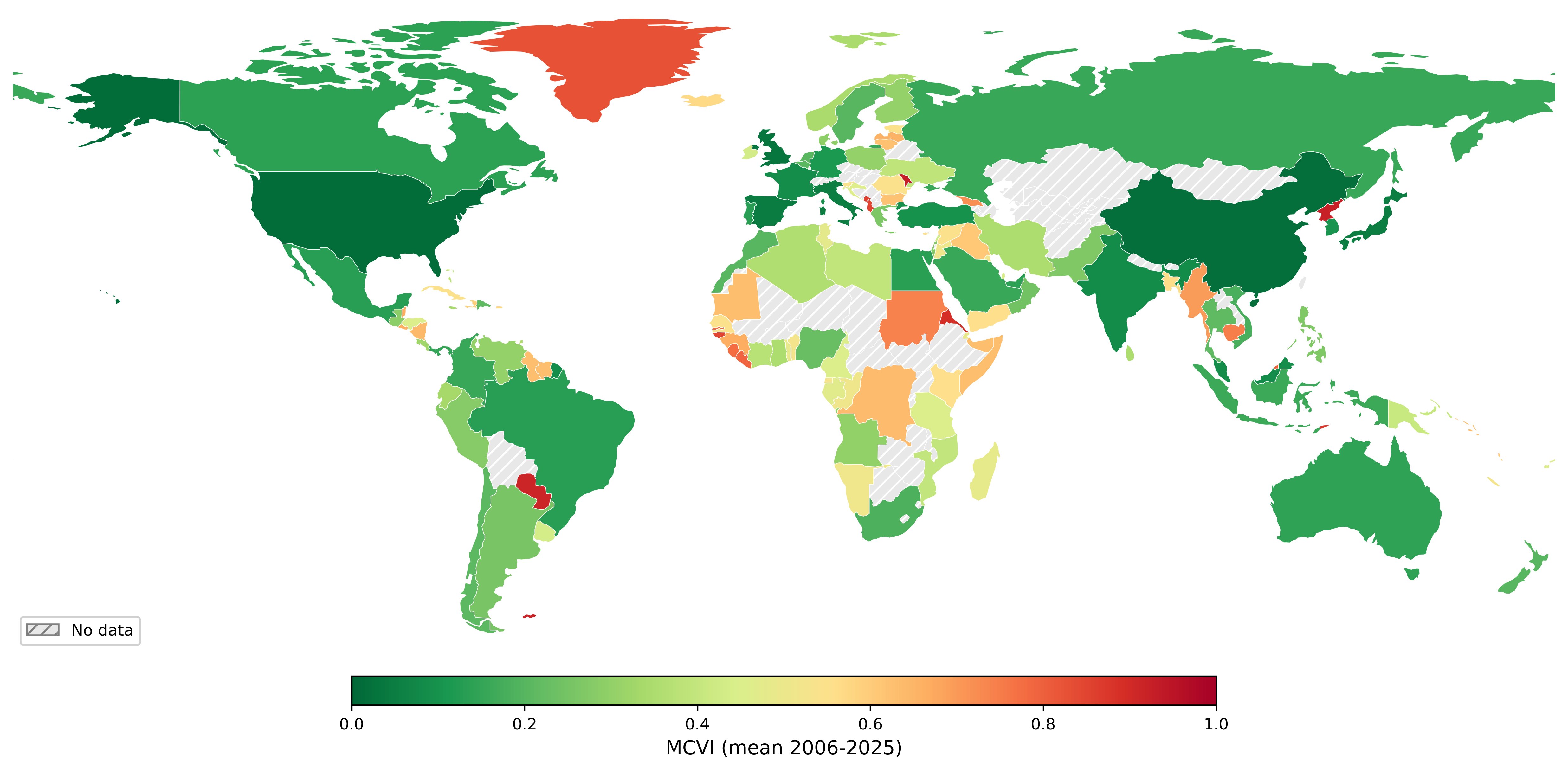}
\caption{Global distribution of the MCVI (mean 2006--2025). Darker red indicates higher vulnerability. Hatched areas indicate no data.}\label{fig:choropleth}
\end{figure}

\Cref{fig:choropleth} maps the global distribution of the MCVI. A clear vulnerability gradient emerges, with low scores concentrated in Europe, East Asia, and North America, and high scores in the Pacific, sub-Saharan Africa, and landlocked regions.

\Cref{tab:topbottom} presents the ten most and least vulnerable economies based on time-averaged MCVI scores.

\begin{table}[H]
\centering
\caption{Top~10 most and least vulnerable economies (mean MCVI, 2006--2025).}\label{tab:topbottom}
\begin{tabular}{l S[table-format=1.3] l S[table-format=1.3]}
\toprule
{Most vulnerable} & {MCVI} & {Least vulnerable} & {MCVI} \\
\midrule
Republic of Moldova & 0.929 & United States & 0.015 \\
Dem.\ People's Rep.\ of Korea & 0.924 & China & 0.019 \\
Saint Helena & 0.923 & United Kingdom & 0.034 \\
Norfolk Island & 0.923 & Spain & 0.044 \\
Falkland Islands & 0.920 & Japan & 0.051 \\
Cocos (Keeling) Islands & 0.919 & Italy & 0.053 \\
Nauru & 0.915 & India & 0.085 \\
Paraguay & 0.912 & France & 0.088 \\
Niue & 0.909 & Malaysia & 0.090 \\
Christmas Island & 0.907 & Turkiye & 0.101 \\
\bottomrule
\end{tabular}
\end{table}

\begin{figure}[H]
\centering
\includegraphics[width=\textwidth]{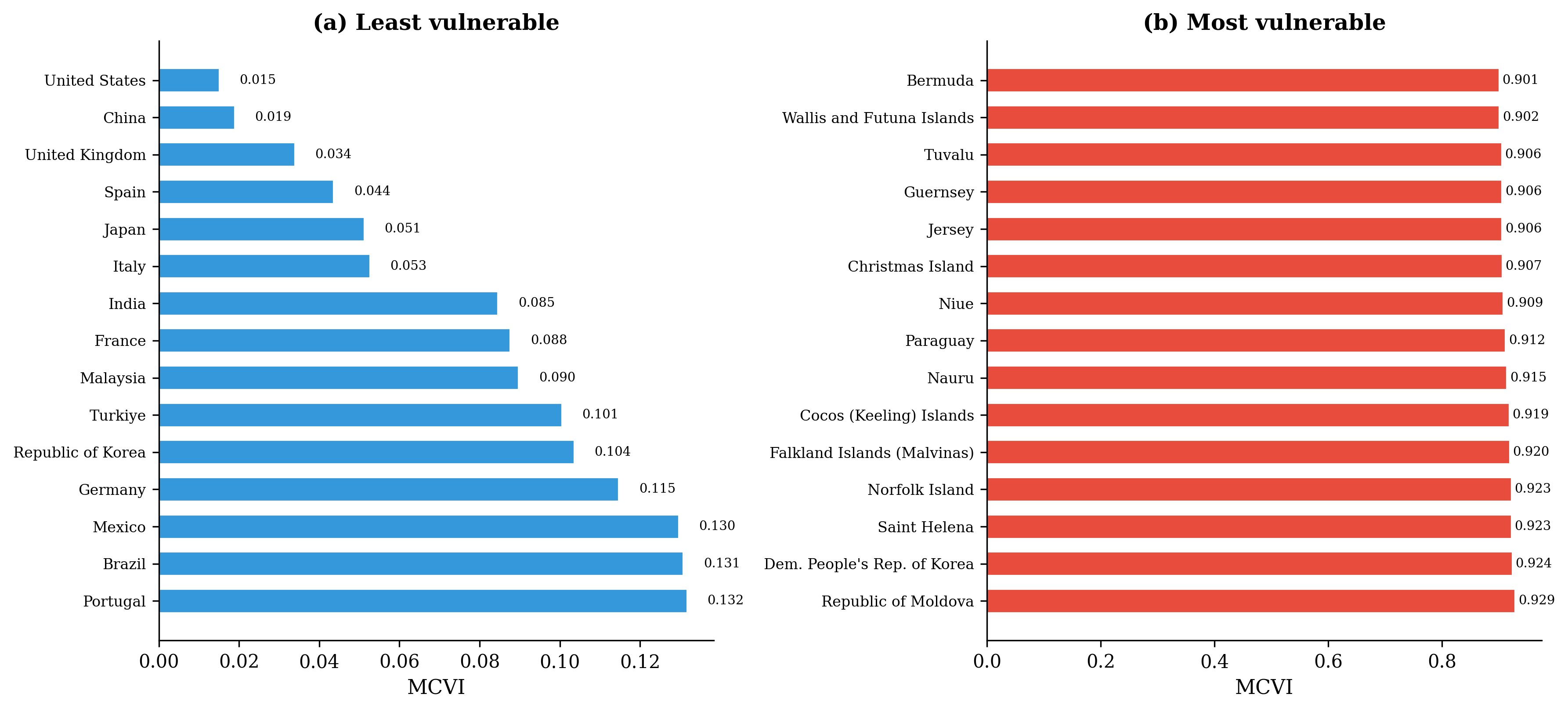}
\caption{The 15 most and least vulnerable economies (mean MCVI, 2006--2025).}\label{fig:topbottom}
\end{figure}

The most vulnerable economies are predominantly small, remote, or landlocked: the Republic of Moldova, classified as a landlocked developing country (LLDC) with minimal direct maritime access, exhibits the highest mean vulnerability. The least vulnerable economies are global maritime hubs with diversified port systems, extensive bilateral networks, and dominant LSCI positions. \Cref{fig:topbottom} displays these rankings graphically. Complete MCVI scores for all 185~economies, including dimension-level components and temporal volatility statistics, are reported in \Cref{tab:appendix} of the appendix.

\begin{figure}[H]
\centering
\includegraphics[width=0.75\textwidth]{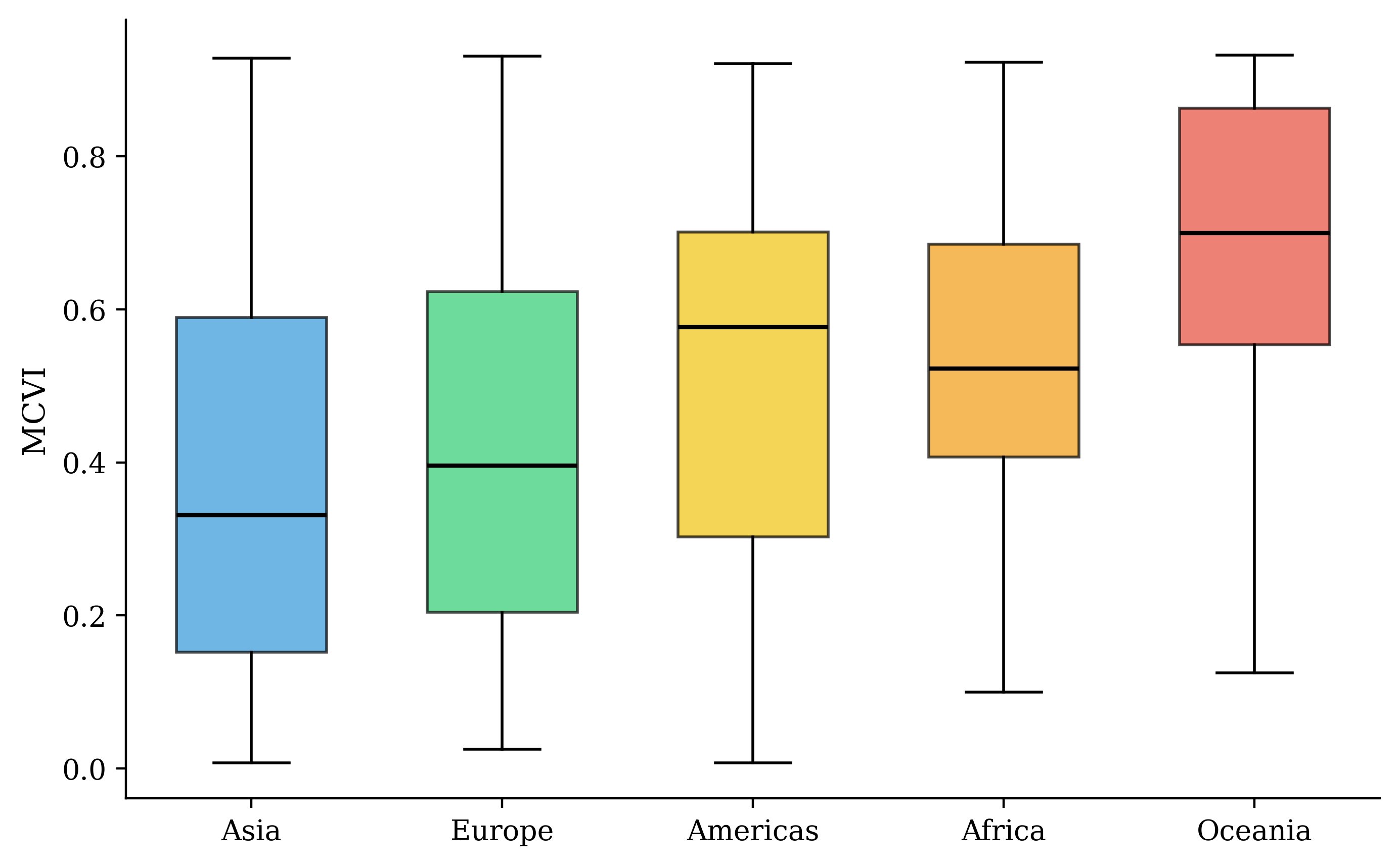}
\caption{MCVI distribution by geographic region.}\label{fig:boxplot}
\end{figure}

Systematic disparities emerge across country classifications (\Cref{tab:groups}). SIDS exhibit a mean MCVI of 0.667 versus 0.433 for non-SIDS economies, a gap of 0.234~points consistent with the connectivity disadvantages documented by \citet{unctad2024rmt} and \citet{rojon2021carbon}. LDCs (0.657) are significantly more vulnerable than non-LDCs (0.471). LLDCs represent the extreme case with a mean MCVI of 0.918, driven by their structural absence of direct maritime access. Regionally, Asian economies exhibit the lowest mean vulnerability (0.393), followed by Europe (0.429), while Oceania---dominated by Pacific island states---displays the highest regional vulnerability (0.669). \Cref{fig:boxplot} visualizes these regional distributions.

\begin{table}[H]
\centering
\caption{Mean MCVI by country classification.}\label{tab:groups}
\begin{tabular}{l S[table-format=1.3] S[table-format=1.3] r}
\toprule
{Group} & {Mean} & {Std.\ dev.} & {$N$ (obs.)} \\
\midrule
Non-SIDS & 0.433 & 0.251 & 2\,481 \\
SIDS & 0.667 & 0.176 & 995 \\
\addlinespace
Non-LDC & 0.471 & 0.258 & 2\,934 \\
LDC & 0.657 & 0.167 & 542 \\
\addlinespace
Non-LLDC & 0.497 & 0.254 & 3\,448 \\
LLDC & 0.918 & 0.011 & 28 \\
\addlinespace
Asia & 0.393 & 0.260 & 667 \\
Europe & 0.429 & 0.273 & 660 \\
Americas & 0.516 & 0.248 & 921 \\
Africa & 0.536 & 0.195 & 790 \\
Oceania & 0.669 & 0.219 & 438 \\
\bottomrule
\end{tabular}
\end{table}

\subsection{Temporal Dynamics (RQ3)}\label{sec:rq3}

\begin{figure}[H]
\centering
\includegraphics[width=0.85\textwidth]{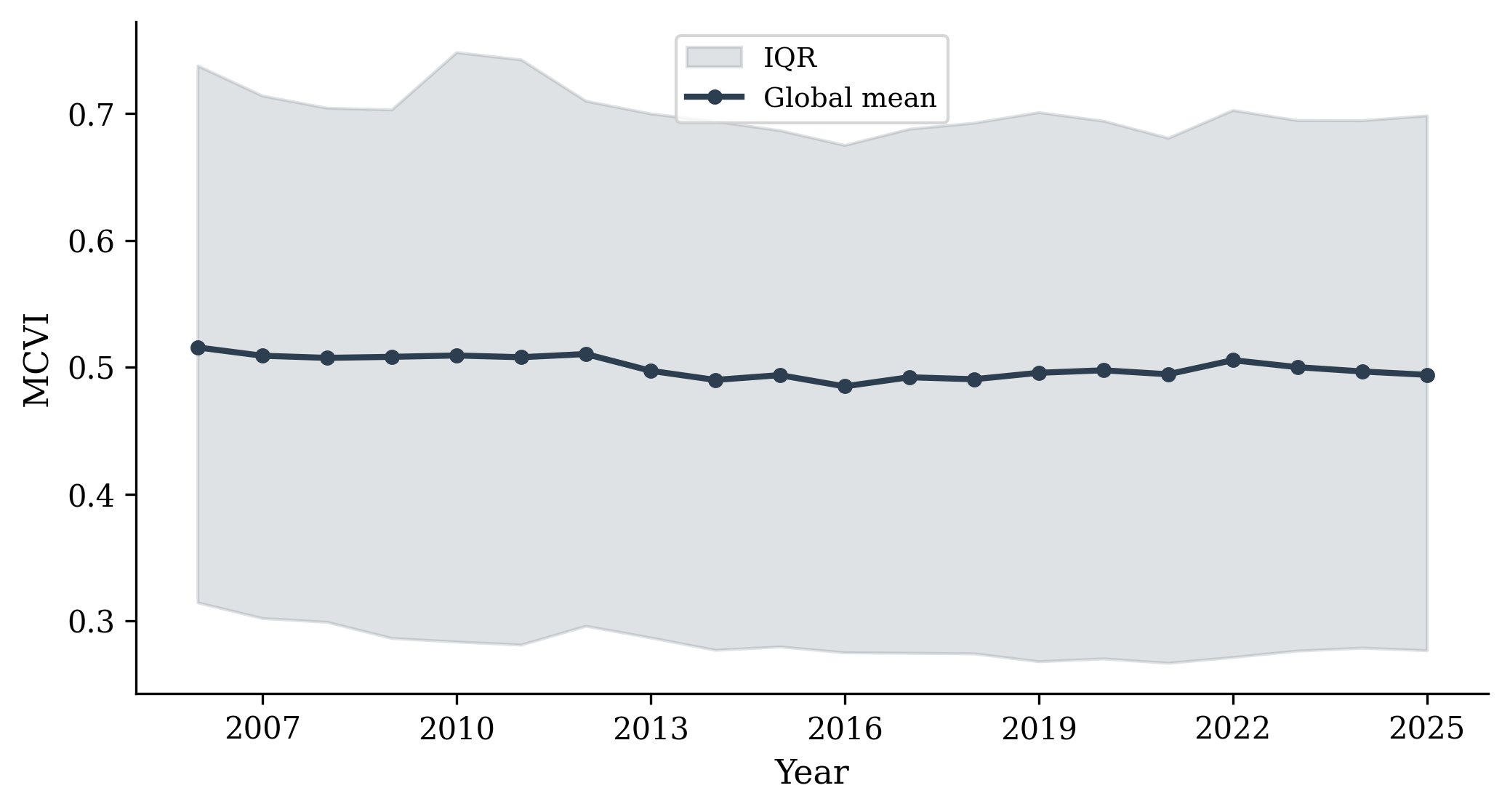}
\caption{Global mean MCVI, 2006--2025, with interquartile range.}\label{fig:globaltrend}
\end{figure}

The global mean MCVI declined from 0.516 in 2006 to 0.494 in 2025, representing a cumulative reduction of $\approx$4.2\% (linear slope: $-0.00087$ per year). This modest downward trend (\Cref{fig:globaltrend}) is consistent with the general expansion of liner shipping capacity and network coverage over the period \citep{unctad2024rmt}. However, the improvement is unevenly distributed.

\begin{figure}[H]
\centering
\includegraphics[width=0.85\textwidth]{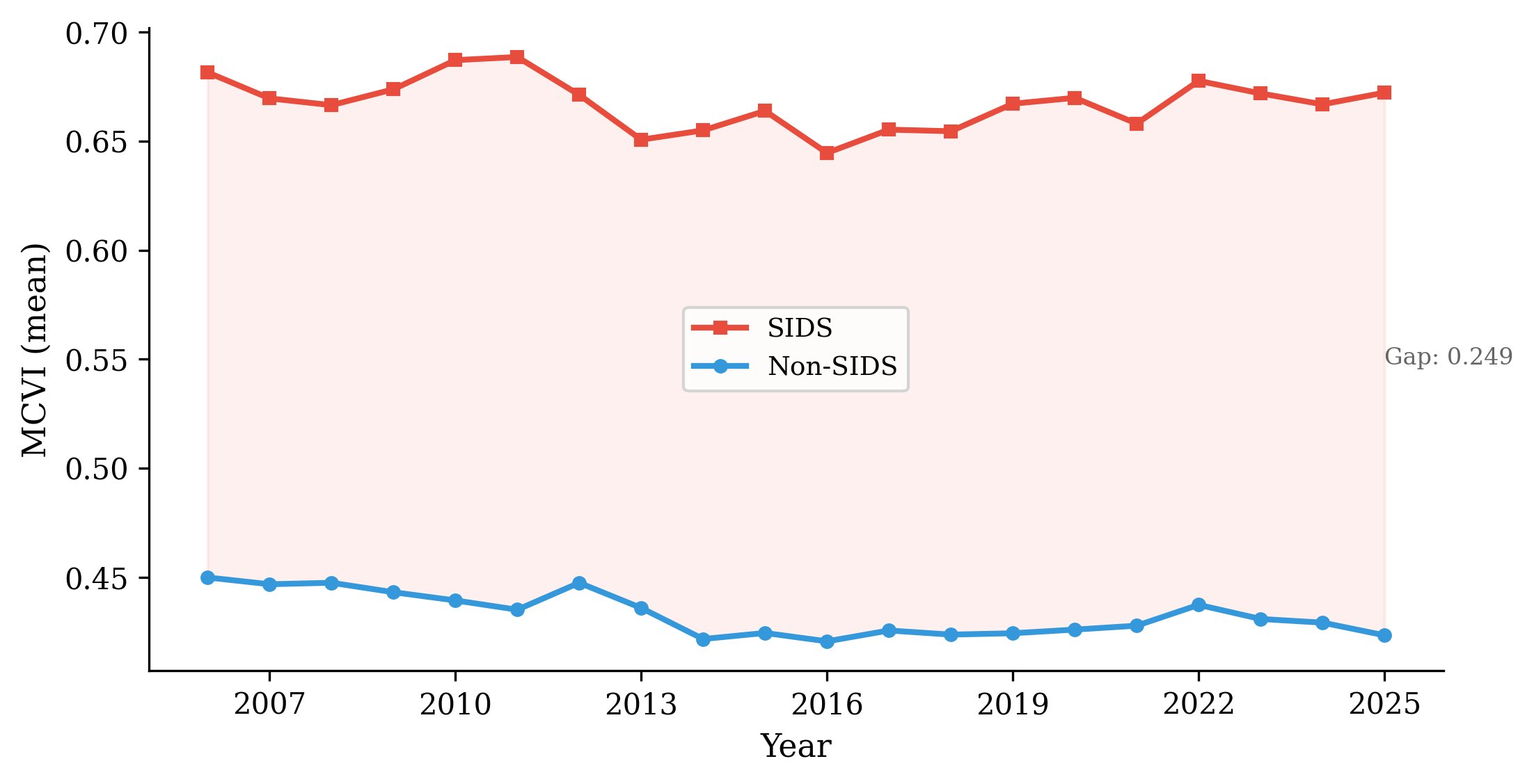}
\caption{MCVI trends for SIDS vs.\ non-SIDS economies, 2006--2025.}\label{fig:sidstrend}
\end{figure}

The SIDS--non-SIDS gap widened from 0.232 in 2006 to 0.249 in 2025 (\Cref{fig:sidstrend}), suggesting that peripheral economies have not benefited proportionally from global connectivity gains. This finding corroborates the observation by \citet{rojon2021carbon} that transport cost reductions have been concentrated among well-connected hub economies.

\begin{figure}[H]
\centering
\includegraphics[width=0.85\textwidth]{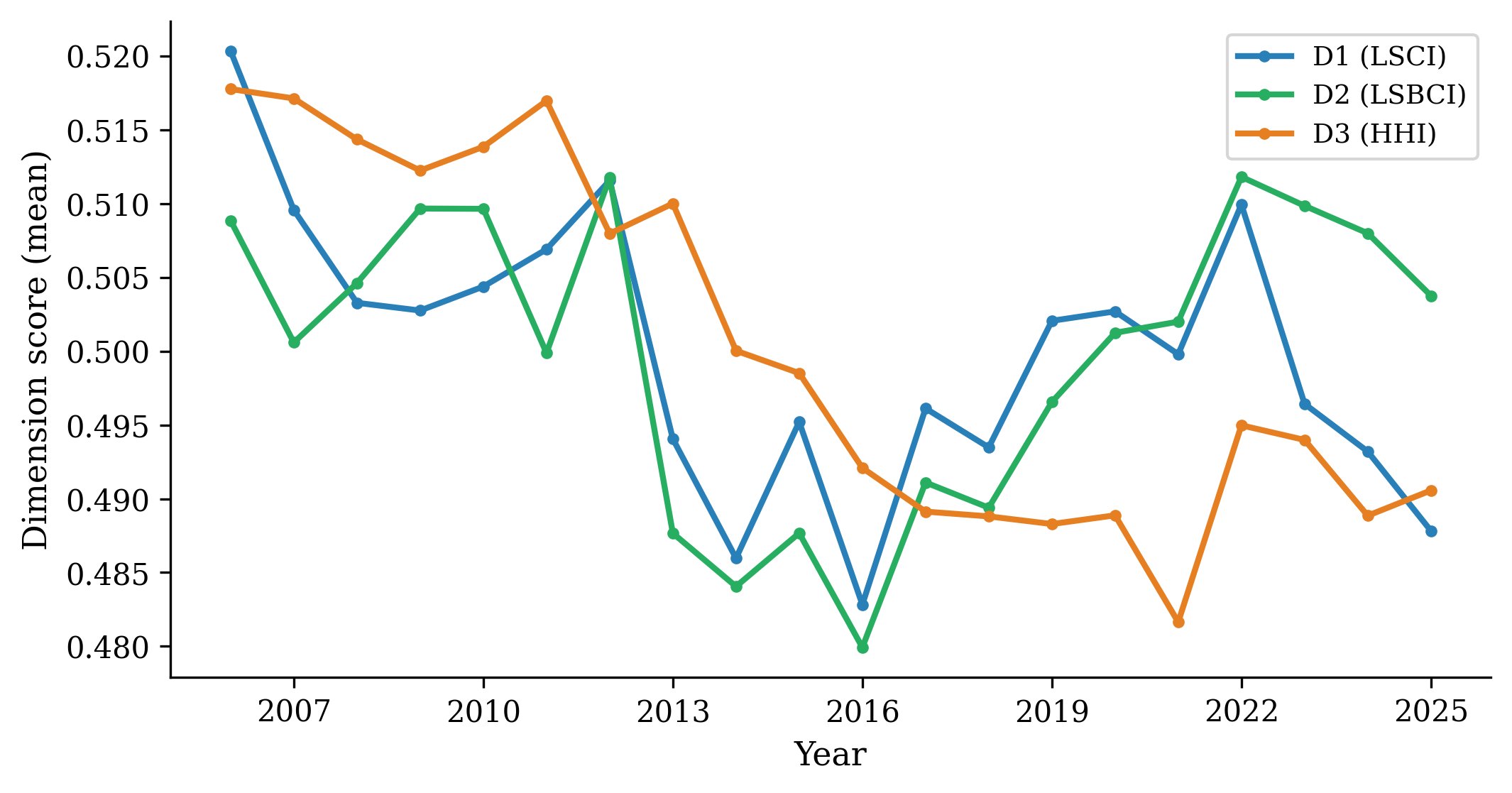}
\caption{Mean dimension scores over time: D1 (LSCI), D2 (LSBCI), D3 (HHI).}\label{fig:dimtrends}
\end{figure}

\Cref{fig:dimtrends} decomposes the global trend by dimension. D1 (connectivity) and D2 (bilateral integration) exhibit similar declining trajectories, consistent with their high correlation. D3 (port concentration) shows a distinct pattern, declining less rapidly, suggesting that infrastructure diversification has lagged behind connectivity improvements.

Rank-order stability between consecutive years is high, with Spearman~$\rho$ ranging from 0.983 to 0.995, indicating that while marginal improvements occur, the structural hierarchy of maritime vulnerability is persistent. Country-level volatility analysis identifies Iraq ($\sigma = 0.156$), Yemen ($\sigma = 0.135$), and Ukraine ($\sigma = 0.127$) as the most variable economies, reflecting geopolitical disruptions that fundamentally altered shipping access. By contrast, major hub economies exhibit near-zero volatility.

\subsection{Decomposition and Clustering (RQ4)}\label{sec:rq4}

\begin{figure}[H]
\centering
\includegraphics[width=0.8\textwidth]{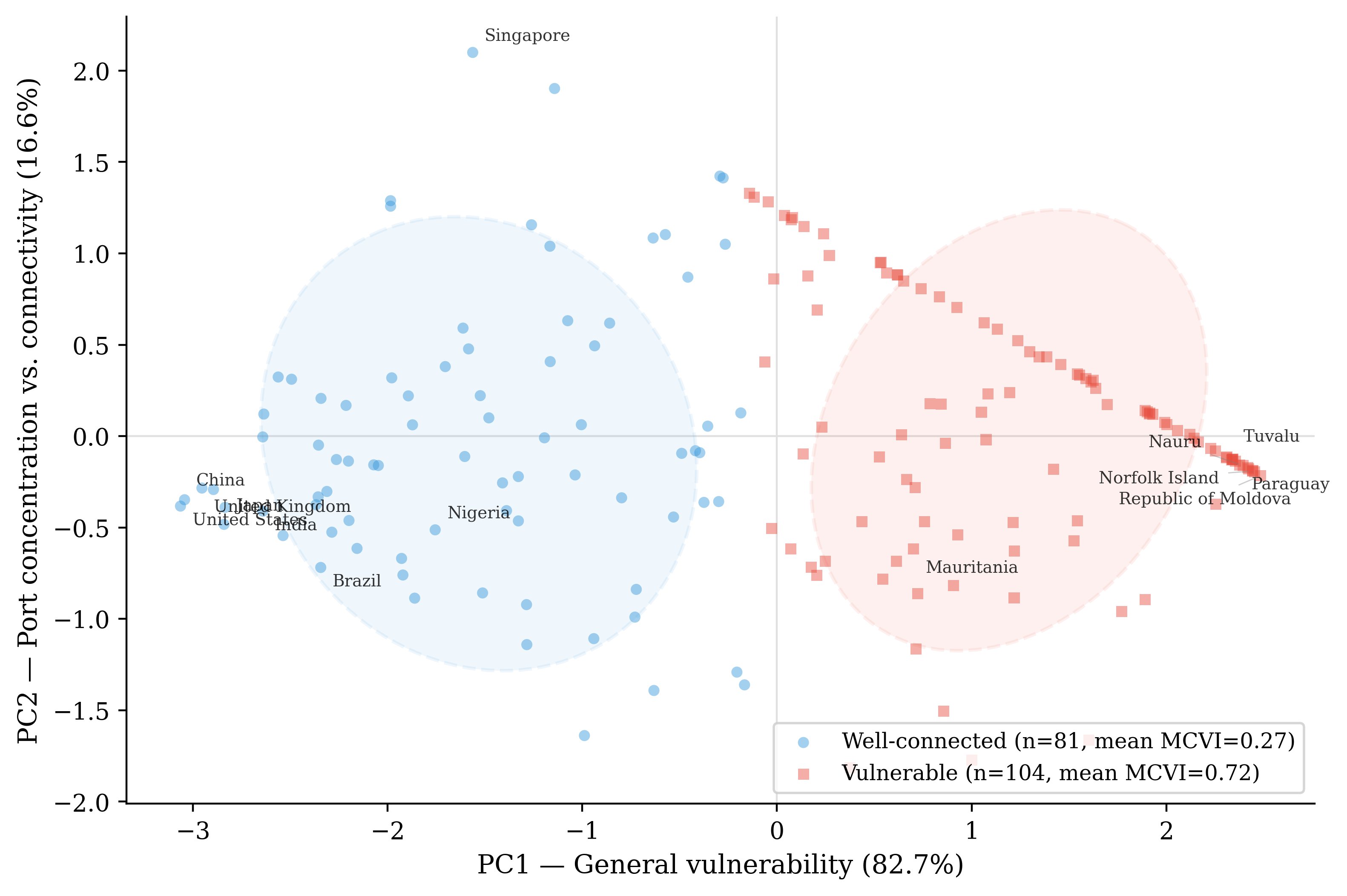}
\caption{PCA scatter plot with $K$-means clusters ($k=2$), computed on country-averaged scores ($n=185$). PC1 captures general vulnerability (82.7\%), PC2 opposes port concentration to connectivity (16.6\%).}\label{fig:pcabiplot}
\end{figure}

$K$-means clustering on time-averaged dimension scores identifies two stable groups (silhouette coefficient~=~0.50). Cluster~1 (\enquote{well-connected}, 81~countries, mean MCVI~$\approx$~0.27) includes most OECD economies and major emerging markets. Cluster~2 (\enquote{vulnerable}, 104~countries, mean MCVI~$\approx$~0.72) encompasses SIDS, LDCs, and small economies. \Cref{fig:pcabiplot} visualizes the two clusters in PCA space.

Dimension decomposition reveals that D3 (port concentration) is the dominant vulnerability source for 72 countries (39\%), D1 (connectivity) dominates for 62 countries (34\%), and D2 (bilateral integration) for 51 countries (27\%). This finding has direct policy implications: for D3-dominant economies, diversifying port infrastructure may yield greater vulnerability reduction than increasing shipping service frequency. The prevalence of D3 as the dominant dimension is consistent with the port criticality findings of \citet{verschuur2021criticality}, who documented that low-income countries are 1.5 times more reliant on their dominant port.

\subsection{Robustness and External Validation (RQ5)}\label{sec:robustness}

\Cref{tab:robustness} summarizes the robustness analysis.

\begin{table}[H]
\centering
\caption{Robustness of MCVI rankings under alternative specifications.}\label{tab:robustness}
\begin{tabular}{l S[table-format=1.4]}
\toprule
{Specification} & {Spearman $\rho$} \\
\midrule
PCA-derived weights (0.36, 0.35, 0.29) & 0.9988 \\
Leave-one-out: drop D1 (LSCI) & 0.9838 \\
Leave-one-out: drop D2 (LSBCI) & 0.9766 \\
Leave-one-out: drop D3 (HHI) & 0.9544 \\
Within-year rank normalization & 0.9988 \\
Pooled min-max normalization & 0.9708 \\
\bottomrule
\end{tabular}
\end{table}

PCA-derived weights (0.362, 0.353, 0.285) produce rankings nearly identical to equal weights ($\rho = 0.999$). All leave-one-dimension-out variants correlate at $\rho > 0.95$, with D3 removal producing the largest deviation ($\rho = 0.954$), confirming that port concentration contributes the most unique information to the index. Alternative normalization methods (within-year ranks, pooled min-max without winsorization) yield $\rho > 0.97$, demonstrating methodological stability.

\subsubsection{Monte Carlo Uncertainty Analysis}\label{sec:montecarlo}

The Monte Carlo simulation simultaneously perturbs aggregation weights, indicator values, and normalization method across 1{,}000~replications \citep{oecd2008handbook, saisana2005}. \Cref{tab:montecarlo} summarizes the results.

\begin{table}[H]
\centering
\caption{Monte Carlo uncertainty analysis: summary statistics (1{,}000 simulations).}\label{tab:montecarlo}
\begin{tabular}{l r}
\toprule
{Metric} & {Value} \\
\midrule
Mean Spearman $\rho$ vs.\ baseline & 0.997 \\
Minimum $\rho$ across simulations & 0.983 \\
\% of simulations with $\rho > 0.95$ & 100.0\% \\
\% of simulations with $\rho > 0.99$ & 98.0\% \\
\addlinespace
Mean rank 95\% CI width & 12.2 positions \\
Median rank 95\% CI width & 11.0 positions \\
Maximum rank 95\% CI width & 37.0 positions \\
\addlinespace
\multicolumn{2}{l}{\textit{Variance decomposition (Sobol-like)}} \\
\quad Weight perturbation & 83.0\% \\
\quad Data noise ($\pm 5\%$) & 16.3\% \\
\quad Normalization choice & 0.7\% \\
\bottomrule
\end{tabular}
\end{table}

Rankings are highly stable: the mean Spearman correlation between simulated and baseline rankings is $\rho = 0.997$, and all 1{,}000~simulations exceed $\rho = 0.95$. The average 95\% confidence interval on a country's rank spans 12.2~positions out of 185---approximately $\pm 6$ ranks---indicating that most countries occupy a narrow and well-determined band in the vulnerability ranking.

Rank uncertainty is concentrated in the middle of the distribution, where countries with similar MCVI scores are more easily permuted by perturbations. At the extremes, rankings are near-deterministic: seven economies (United States, China, United Kingdom, Spain, Japan, Italy, France) remain in the bottom~10 in 100\% of simulations, and the two most vulnerable economies with full panel coverage (Norfolk Island, Bermuda) remain in the top~10 in over 96\% of simulations. Singapore exhibits the widest confidence interval among well-connected economies (rank $43 \pm 15$), reflecting the tension between its excellent connectivity (low D1) and single-port structure (high D3)---a substantively meaningful sensitivity rather than a methodological artifact.

The Sobol-like variance decomposition \citep{saltelli2008} reveals that 83\% of total rank uncertainty is attributable to weight perturbation, 16\% to data noise, and less than 1\% to the normalization choice. The dominance of weight uncertainty is expected given three dimensions: even moderate Dirichlet perturbations around $(1/3, 1/3, 1/3)$ alter the relative importance of D3 (port concentration) versus D1--D2 (connectivity--bilateral integration). Critically, despite the high D1--D2 correlation ($\rho = 0.964$), which exceeds the JRC threshold of 0.95 for potential dimension domination \citep{oecd2008handbook}, the Monte Carlo analysis demonstrates that this correlation does not compromise ranking stability: simultaneous perturbation of all uncertainty sources still produces $\rho > 0.95$ in every simulation. This result provides a stronger defense of the three-dimension architecture than the deterministic leave-one-out tests alone, as it accounts for the joint effect of multiple perturbations rather than isolated parameter changes \citep{saisana2005, greco2018}.

\subsubsection{Temporal Robustness}\label{sec:temporal_robust}

To assess whether rankings are stable over time, split-half and predictive tests are conducted. The Spearman correlation between mean MCVI computed over 2006--2015 and mean MCVI over 2016--2025 is $\rho = 0.970$ ($n = 180$, $p < 10^{-110}$). The correlation between single-year rankings in 2006 and 2025 is $\rho = 0.929$ ($n = 164$), indicating high predictive stability over the full 20-year span. Five-year rolling correlations between adjacent periods (2006--2010 vs.\ 2011--2015, etc.) all exceed $\rho = 0.978$. These results confirm that the structural hierarchy of maritime vulnerability is persistent and that the MCVI is not an artifact of short-term fluctuations.

\subsubsection{Convergent Validity: LPI and Transport Costs}\label{sec:convergent}

Two external indicators are used for convergent validity testing. The World Bank Logistics Performance Index (LPI), available for seven waves between 2007 and 2022, provides a broad measure of trade logistics efficiency \citep{worldbank2023lpi}. The UNCTAD--World Bank Trade-and-Transport Dataset provides country-level ad valorem maritime freight rates for 2016--2021 \citep{unctadtransport2024}.

\Cref{tab:convergent} reports the results. The MCVI correlates negatively with LPI across all seven waves (average $\rho = -0.61$), confirming that countries with higher maritime vulnerability systematically exhibit lower logistics performance. The correlation with maritime ad valorem freight rates is positive across all six years (average $\rho = +0.32$, all $p < 0.001$), indicating that more vulnerable countries pay higher freight costs relative to the value of their imports, as documented by \citet{rojon2021carbon} for SIDS and LDCs.

\begin{table}[H]
\centering
\caption{Convergent validity: MCVI correlations with external indicators.}\label{tab:convergent}
\begin{tabular}{l l S[table-format=-1.4] S[table-format=3.0]}
\toprule
{Indicator} & {Year} & {Spearman $\rho$} & {$n$} \\
\midrule
\multirow{7}{*}{LPI (World Bank)} & 2007 & -0.6344 & 106 \\
& 2010 & -0.5981 & 112 \\
& 2012 & -0.6572 & 113 \\
& 2014 & -0.5746 & 113 \\
& 2016 & -0.6054 & 117 \\
& 2018 & -0.5919 & 117 \\
& 2022 & -0.6008 & 105 \\
\addlinespace
\multirow{6}{*}{Ad valorem freight rate (UNCTAD)} & 2016 & 0.3085 & 124 \\
& 2017 & 0.2878 & 125 \\
& 2018 & 0.3535 & 125 \\
& 2019 & 0.3192 & 125 \\
& 2020 & 0.3106 & 125 \\
& 2021 & 0.3357 & 125 \\
\bottomrule
\multicolumn{4}{l}{\footnotesize All correlations significant at $p < 0.001$.}
\end{tabular}
\end{table}

The stronger LPI correlation ($|\rho| \approx 0.61$) compared with transport costs ($\rho \approx 0.32$) is expected: the LPI captures infrastructure and institutional dimensions that are structurally correlated with connectivity, whereas ad valorem freight rates also depend on commodity composition, trade imbalances, and distance---factors not captured by the MCVI.

\subsubsection{External Validation: Panel Regressions}\label{sec:regression}

\begin{figure}[H]
\centering
\includegraphics[width=0.7\textwidth]{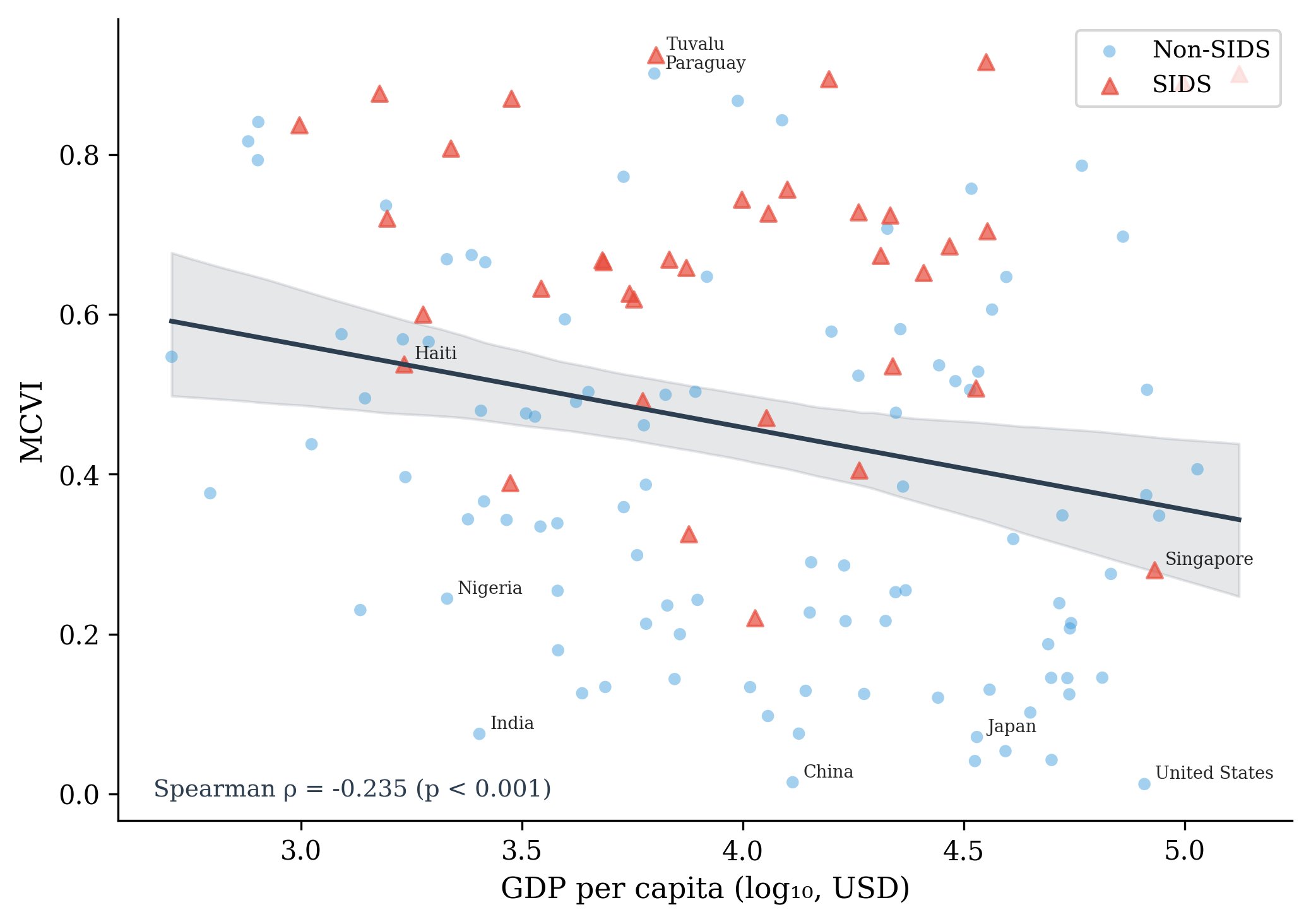}
\caption{MCVI vs.\ GDP per capita (2023) with OLS regression line and 95\% confidence interval. SIDS economies (triangles) cluster in the high-vulnerability, low-GDP quadrant.}\label{fig:scatter}
\end{figure}

To assess external validity further, pooled OLS regressions with standard errors clustered at the country level are estimated on a subsample of 2,324~observations (130~countries, 2006--2024) matched with GDP per capita from the World Bank and trade openness (trade-to-GDP ratio). \Cref{tab:regression} reports the results and \Cref{fig:scatter} illustrates the cross-sectional relationship between MCVI and GDP per capita, distinguishing SIDS from non-SIDS economies.

\begin{table}[H]
\centering
\caption{Panel regression results: MCVI as dependent variable.}\label{tab:regression}
\begin{tabular}{l S[table-format=-1.4] S[table-format=-1.4] S[table-format=-1.4]}
\toprule
& {Model 1} & {Model 2} & {Model 3} \\
\midrule
Constant & 0.9063 & -0.0460 & 0.2266 \\
& {(7.10)} & {(0.23)} & {(1.12)} \\
$\log(\text{GDP/cap})$ & -0.0510 & & -0.0432 \\
& {(3.54)} & & {(2.72)} \\
$\log(\text{Trade/GDP})$ & & 0.1131 & 0.1251 \\
& & {(2.53)} & {(3.35)} \\
SIDS dummy & & & 0.2301 \\
& & & {(6.17)} \\
LDC dummy & & & 0.1060 \\
& & & {(2.10)} \\
\addlinespace
$R^2$ & 0.078 & 0.057 & 0.350 \\
$N$ & {2{,}324} & {2{,}324} & {2{,}324} \\
\bottomrule
\multicolumn{4}{l}{\footnotesize Absolute $t$-statistics in parentheses (clustered SE at the country level).}\\
\multicolumn{4}{l}{\footnotesize Main coefficients significant at $p<0.05$ or better; $\log(\text{GDP/cap})$ and SIDS dummy at $p<0.01$.}
\end{tabular}
\end{table}

Model~1 confirms a negative association between MCVI and GDP per capita ($\beta = -0.051$, $p < 0.001$): richer countries tend to be less vulnerable, consistent with the established relationship between economic development and maritime connectivity \citep{fugazza2017, hoffmann2020}.

The positive coefficient on trade openness in Model~2 ($\beta = 0.113$, $p < 0.05$) reveals a vulnerability paradox: more trade-open economies tend to exhibit \emph{higher} vulnerability. This reflects the composition of the group---small, trade-dependent economies (especially SIDS) have high trade-to-GDP ratios precisely because they import most goods by sea, yet lack the scale economies and geographic advantages that reduce connectivity costs \citep{rojon2021carbon}. This finding cautions against interpreting MCVI purely as a policy performance indicator; the index captures structural vulnerability as well as policy-amenable factors.

Model~3 adds SIDS and LDC dummies, raising $R^2$ to 0.350. SIDS status contributes 0.230~additional MCVI points (significant at $p < 0.001$), and LDC status contributes 0.106~points (significant at $p < 0.05$), underscoring the structural nature of vulnerability for these country groups. The $R^2$ values reflect the modest explanatory power of macroeconomic variables alone; the remaining variation captures structural factors (geography, historical trade patterns, port infrastructure) beyond the specified controls. Importantly, a modest $R^2$ in this context is a feature rather than a limitation: if standard macroeconomic regressors explained most of the MCVI variance, the index would be statistically redundant with these variables and would add little analytical value beyond them. The fact that the MCVI captures substantial structural variation \emph{orthogonal} to GDP per capita and trade openness is precisely what justifies its construction as an independent diagnostic instrument for cross-country vulnerability assessment.

As a sensitivity check, fixed-effects (FE) and random-effects (RE) panel specifications are estimated. The Hausman test strongly favors FE over RE ($H = 370.76$, $p < 0.001$). Under FE, which absorbs time-invariant characteristics including SIDS and LDC status, log GDP per capita retains its negative association with MCVI ($\beta = -0.052$, $t = -5.68$, $p < 0.001$), indicating that within-country economic growth is associated with modest vulnerability reduction. However, trade openness becomes insignificant under FE ($\beta = -0.009$, $t = -0.73$), and the within-$R^2$ is only 0.071. This low within-$R^2$ is substantively informative: it confirms that cross-sectional variation between countries dominates temporal variation within countries, consistent with the high rank stability documented in \Cref{sec:temporal_robust} and reinforcing the interpretation of the MCVI as a measure of persistent structural vulnerability rather than a cyclical indicator.

\subsection{Predictive Validity: Disruption Event Analysis}\label{sec:events}

A composite vulnerability index should predict realized impacts during disruption events. Two contrasting shocks provide natural experiments: the COVID-19 pandemic (2020), a supply-side shock that disrupted port operations, shipping schedules, and container logistics globally \citep{notteboom2021, guerrero2022covid}; and the 2008--2009 global financial crisis, a demand-side shock that contracted trade volumes through reduced consumption and investment \citep{notteboom2021}. The Red Sea crisis (2023--2024), a route-specific disruption caused by Houthi attacks on commercial shipping \citep{redsea2024mel}, is examined as a supplementary case.

\subsubsection{COVID-19 Supply Shock (2019--2020)}

Countries are ranked by their 2019 MCVI and divided into quartiles. \Cref{tab:events} reports trade openness changes by vulnerability group. A clear gradient emerges: the least vulnerable quartile (Q1) experienced a mean trade decline of $-7.3\%$, while the most vulnerable quartiles (Q3 and Q4) experienced $-13.0\%$ and $-12.3\%$, respectively. The Spearman correlation between pre-crisis MCVI and trade change is $\rho = -0.251$ ($p = 0.004$, $n = 127$), indicating that structurally vulnerable countries suffered significantly larger trade losses. A Mann-Whitney $U$-test confirms that Q4 experienced statistically worse outcomes than Q1 ($p = 0.013$).

The SIDS--non-SIDS comparison reinforces this finding: SIDS economies lost $-14.1\%$ of trade openness compared with $-9.4\%$ for non-SIDS ($p = 0.021$). Countries with concentrated port infrastructure, few bilateral partners, and low liner connectivity had fewer alternatives when shipping services were disrupted, amplifying the transmission of the supply shock to their trade performance \citep{verschuur2021criticality}.

\begin{table}[H]
\centering
\caption{Trade impact of disruption events by pre-crisis MCVI quartile.}\label{tab:events}
\begin{tabular}{l S[table-format=-2.1] S[table-format=-2.1]}
\toprule
& \multicolumn{2}{c}{$\Delta$Trade openness (\%)} \\
\cmidrule(l){2-3}
{MCVI quartile} & {COVID-19 ($n=127$)} & {Fin.\ crisis ($n=120$)} \\
\midrule
Q1 (least vulnerable) & -7.3 & -13.4 \\
Q2 & -9.4 & -11.2 \\
Q3 & -13.0 & -11.0 \\
Q4 (most vulnerable) & -12.3 & -7.5 \\
\addlinespace
{Spearman $\rho$ (MCVI vs.\ $\Delta$Trade)} & -0.251 & {+0.233} \\
{$p$-value} & {0.004} & {0.010} \\
\bottomrule
\end{tabular}
\end{table}

\subsubsection{Financial Crisis Demand Shock (2008--2009)}

The 2008--2009 financial crisis produced a strikingly different pattern. The Spearman correlation between pre-crisis MCVI and trade change is $\rho = +0.233$ ($p = 0.010$, $n = 120$)---positive rather than negative. Well-connected countries (Q1) lost $-13.4\%$ of trade, while the most vulnerable (Q4) lost only $-7.5\%$. This reversal is economically coherent: the financial crisis was a demand contraction centered on advanced economies. Countries deeply integrated into global supply chains---precisely those with low MCVI scores---experienced the largest trade declines because their export volumes were most sensitive to falling consumer demand in major markets \citep{notteboom2021}. Vulnerable economies, which trade less and are less integrated into global value chains, were partially insulated from the demand shock.

The opposing signs across the two events ($\rho = -0.25$ for the supply shock, $\rho = +0.23$ for the demand shock) constitute a discriminant validity test: the MCVI predicts vulnerability specifically to supply-side maritime disruptions, not to trade shocks in general. This specificity is consistent with the conceptual definition in \Cref{sec:definition}, which scopes vulnerability to the supply side of liner shipping connectivity.

\subsubsection{Red Sea Route Disruption (2023--2024)}

The Red Sea crisis, triggered by Houthi attacks on commercial shipping from late 2023, led to widespread rerouting of Asia--Europe traffic via the Cape of Good Hope \citep{redsea2024mel, unctad2024rmt}. Unlike the COVID-19 pandemic, this disruption was route-specific rather than systemic. The Spearman correlation between 2023 MCVI and trade change is $\rho = -0.094$ ($p = 0.33$, $n = 108$)---not statistically significant. This null result is informative: it confirms that the MCVI, which captures structural connectivity characteristics, does not predict the impact of chokepoint-specific disruptions. Vulnerability to route disruptions depends on trade corridor exposure, commodity composition, and carrier rerouting decisions---factors beyond the scope of a supply-side connectivity index. This result supports the limitation noted in \Cref{sec:definition} and suggests that incorporating chokepoint exposure indicators would complement the MCVI framework in future extensions.

\section{Discussion}\label{sec:discussion}

\subsection{Principal Findings and Their Significance}

The MCVI provides the first composite measure of supply-side maritime connectivity vulnerability at the country level with consistent temporal coverage from 2006 to 2025. The results yield six principal insights that advance the understanding of structural inequalities in global shipping networks.

\textit{The connectivity divide is structural and persistent.} The SIDS premium of 0.234 MCVI points represents a gap that persisted---and widened marginally from 0.232 to 0.249---over two decades of global maritime expansion. This finding is consistent with but extends beyond the qualitative observations of \citet{unctad2024rmt}, who reported that SIDS are on average ten times less connected than non-SIDS economies. The MCVI quantifies this disadvantage as a composite structural condition rather than a single-dimension deficit and demonstrates that the gap is not narrowing despite decades of trade facilitation efforts. For LLDCs, the extreme mean MCVI of 0.918 underscores that landlocked status translates almost deterministically into maximum maritime vulnerability, a finding that complements the trade cost analyses of \citet{fugazza2017}.

\textit{Port concentration is the primary vulnerability driver for nearly 40\% of economies.} The decomposition analysis reveals that D3 is the dominant dimension for 72 out of 185~countries. This prevalence is consistent with the asset-level risk analysis of \citet{verschuur2023multihazard}, who found that 86\% of ports globally are exposed to multiple hazards, and with the port criticality assessment of \citet{verschuur2021criticality}, who documented that low-income countries are 1.5~times more reliant on their dominant port. The MCVI extends these port-level findings to the country level, showing that single-port dependency is the most discriminating factor in cross-country vulnerability rankings.

\textit{The vulnerability paradox of trade openness.} The positive association between trade openness and MCVI ($\beta = 0.113$, $p < 0.05$) constitutes an original finding. Small, trade-dependent economies---particularly SIDS---exhibit high trade-to-GDP ratios precisely because they import most essential goods by sea, yet their limited market size precludes the scale economies that attract frequent liner services \citep{rojon2021carbon}. This paradox has implications for the interpretation of maritime connectivity as a policy lever: for many vulnerable economies, the constraint is not insufficient trade liberalization but insufficient shipping supply relative to trade needs.

\textit{The MCVI captures supply-side vulnerability specifically.} The disruption event analysis provides discriminant validation. During the COVID-19 supply shock, structurally vulnerable countries (high MCVI) suffered disproportionately larger trade losses ($\rho = -0.25$, $p < 0.005$), consistent with the mechanism that limited routing alternatives amplify supply disruptions \citep{verschuur2021criticality, guerrero2022covid}. During the 2008--2009 demand shock, the relationship reversed ($\rho = +0.23$, $p = 0.01$): well-connected economies, more deeply embedded in global value chains, lost proportionally more trade \citep{notteboom2021}. This opposing pattern confirms that the MCVI captures structural fragility of maritime service supply rather than general trade exposure---a distinction central to the conceptual framework (\Cref{sec:definition}). The null result for the Red Sea route disruption ($\rho = -0.09$, $p = 0.33$) further delineates the scope: the MCVI does not predict impacts from chokepoint-specific events, consistent with its design as a connectivity-based rather than route-based vulnerability measure.

\textit{Global vulnerability has improved only marginally.} The $\approx$4.2\% cumulative decline over 20~years translates to approximately 0.2\% per year---a rate substantially below the growth in global container fleet capacity over the same period. This suggests that fleet expansion has disproportionately benefited already well-connected routes and hubs, consistent with the network concentration dynamics analyzed by \citet{wang2023glsn} and the entropy-based vulnerability model of \citet{ducruet2022entropy}.

\textit{The MCVI is methodologically robust.} Rankings are stable across all tested alternative specifications ($\rho > 0.95$). The PCA-derived weights ($w = 0.36, 0.35, 0.29$) are close to equal weights, validating the baseline aggregation. The drop-D3 test produces the largest ranking deviation ($\rho = 0.954$), confirming that port concentration carries the most unique information---a result with implications for index design in future extensions. Monte Carlo simulation, simultaneously perturbing weights, data, and normalization across 1{,}000 replications, confirms that $\rho > 0.95$ in every simulation and that 83\% of rank uncertainty is attributable to weight choice---a source already shown to have negligible impact on rankings \citep{saisana2005}. Despite the high D1--D2 correlation ($\rho = 0.964$), which exceeds the JRC domination threshold \citep{oecd2008handbook}, ranking stability is not compromised.

\subsection{Policy Implications}

The MCVI has three categories of policy application. First, as a \textit{monitoring tool}, it enables tracking of maritime vulnerability over time and across country groups, supporting the evidence base for SDG~Target~17.11 (increase exports of developing countries) and the SAMOA Pathway objectives for SIDS \citep{unctad2024rmt}. Second, as a \textit{diagnostic instrument}, the dimension decomposition identifies the primary vulnerability channel for each country, enabling targeted interventions---port diversification for D3-dominant economies, bilateral shipping agreements for D2-dominant economies, and general connectivity enhancement for D1-dominant economies. Third, as an \textit{analytical variable}, the MCVI can serve in econometric studies as either a dependent variable (studying determinants of vulnerability reduction) or an independent variable (studying the impact of maritime vulnerability on trade volumes, prices, or supply chain resilience).

\subsection{Limitations}

The MCVI has five principal limitations. First, it relies exclusively on supply-side liner shipping data and does not capture demand-side factors such as trade volumes, commodity composition, or essential import dependency. Second, the high D1--D2 correlation ($\rho = 0.964$) means that the effective dimensionality is closer to two than three, which may limit discriminating power for countries with divergent connectivity and bilateral profiles; however, Monte Carlo simulation demonstrates that this correlation does not compromise ranking stability ($\rho > 0.95$ in 100\% of 1{,}000 simulations). Third, the underlying UNCTAD datasets reflect scheduled services and reported port capacities rather than realized vessel movements (AIS data), and may not fully capture informal shipping services or transshipment-dependent connections prevalent in SIDS \citep{unctad2022asean}. This positions the MCVI as an \emph{ex ante} structural measure, conceptually distinct from and complementary to AIS-based indicators of realized disruption such as the GSCSI-M \citep{arvis2026} and the service-level VTP framework \citep{fan2025vtp}. Fourth, the pooled normalization approach means that country scores are relative to the panel composition; the entry or exit of economies marginally affects all other scores. Fifth, the MCVI does not account for chokepoint exposure or route-specific vulnerability, which would require vessel-tracking data beyond publicly available sources \citep{arvis2026}.

\section{Conclusion}\label{sec:conclusion}

This paper has introduced the Maritime Connectivity Vulnerability Index (MCVI), the first composite indicator measuring supply-side liner shipping vulnerability at the country level with consistent temporal coverage. The index is constructed from three dimensions---national connectivity (LSCI), bilateral integration (LSBCI), and port concentration (PLSCI)---using pooled fractional rank normalization and equal-weight aggregation across 185~economies over the period 2006--2025.

The analysis yields six key findings with implications for maritime economics, trade policy, and development strategy. First, SIDS, LDCs, and LLDCs face systematically higher maritime vulnerability, with a SIDS premium of 0.234~points that has widened over the sample period---indicating that two decades of global maritime expansion have not closed the connectivity divide. Second, the global mean MCVI declined by only 4.2\%, a rate far below the growth in container fleet capacity, suggesting that new shipping supply has been absorbed disproportionately by already well-served routes. Third, port concentration emerges as the dominant vulnerability channel for nearly 40\% of economies, establishing infrastructure diversification---not just connectivity improvement---as a policy priority. Fourth, a vulnerability paradox of trade openness is identified: small, trade-dependent economies are simultaneously the most trade-open and the most maritimely vulnerable, challenging simplistic narratives that equate trade openness with resilience. Fifth, the disruption event analysis demonstrates that the MCVI has predictive validity for supply-side shocks: pre-crisis vulnerability scores predict COVID-19 trade losses ($\rho = -0.25$, $p < 0.005$), while the contrasting positive association during the 2008--2009 demand shock ($\rho = +0.23$) confirms the supply-side specificity of the index. Sixth, the index is robust across all tested specifications, with rank correlations exceeding 0.95 for every alternative weighting, normalization, and dimensional configuration.

The MCVI makes three original contributions to the literature. \textit{Methodologically}, it demonstrates that a parsimonious three-dimension architecture using exclusively public data can produce a robust composite index of maritime vulnerability, complementing existing connectivity metrics \citep{hoffmann2005, wang2022cci, mishra2021} with an explicit vulnerability perspective; the combination of pooled fractional rank normalization and Monte Carlo uncertainty propagation (weights, data, and normalization jointly perturbed) provides a methodological template for vulnerability indicators in contexts with mildly unbalanced panels. \textit{Empirically}, it provides the first systematic quantification of cross-country maritime vulnerability over a 20-year period, offering a baseline for longitudinal monitoring, and demonstrates---through the COVID-19 and financial crisis natural experiments---that the index has predictive validity specifically for supply-side disruptions. \textit{Substantively}, it introduces three novel findings with direct policy implications: (i)~the \emph{vulnerability paradox of trade openness}, whereby the most trade-open small economies are simultaneously the most maritimely vulnerable; (ii)~the \emph{D3-dominance finding}, establishing port concentration as the primary vulnerability channel for nearly 40\% of economies and thus motivating infrastructure diversification as a distinct policy lever; and (iii)~the \emph{supply--demand shock discrimination}, validated by the opposing signs of MCVI--trade correlations during the COVID-19 supply shock ($\rho = -0.25$) and the 2008--2009 demand shock ($\rho = +0.23$).

Several avenues for future research arise from this work. First, the incorporation of transport cost data from the UNCTAD--World Bank Trade-and-Transport Dataset, as its temporal coverage extends beyond the current 2016--2021 window, would add a fourth dimension capturing the economic burden of shipping. Second, disaggregation of vulnerability by commodity type could reveal sector-specific fragilities, particularly for food and energy imports in SIDS. Third, instrumental variable approaches could explore the causal impact of MCVI reductions on trade volumes and economic growth. Fourth, the MCVI could be linked to event-specific disruption data---such as the Red Sea rerouting of 2023--2024 \citep{redsea2024mel} or the Hormuz crisis of 2026 \citep{hormuz2026}---to assess whether structurally vulnerable countries experience disproportionate trade losses during chokepoint disruptions. Finally, the integration of demand-side indicators (essential import shares, food import dependency) would move from a supply-side vulnerability index toward a comprehensive maritime resilience measure.

\section*{Acknowledgments}
The authors gratefully acknowledge UNCTAD for making the LSCI, LSBCI, and PLSCI datasets publicly available through the UNCTADstat Data Centre, which enabled the construction of the MCVI. 

\section*{Data Availability}
All datasets used in this study are publicly available through the UNCTADstat Data Centre (\url{https://unctadstat.unctad.org}). 

\section*{Declarations}

\subsection*{Funding}
Not applicable.

\subsection*{Conflicts of Interest}
The authors declare that there is no conflict of interest.

\subsection*{Consent for Publication}
All authors have reviewed and approved the final version of the manuscript and have provided consent for its publication.

\subsection*{Ethics Approval}
Not applicable.

\clearpage
\appendix
\setcounter{table}{0}
\renewcommand{\thetable}{A\arabic{table}}

\newgeometry{left=5mm,right=5mm,top=8mm,bottom=5mm}
\begin{landscape}
\pagestyle{lscapestyle}
\section{MCVI Scores for 185 Economies}\label{app:scores}
\vspace{-4pt}

\setlength{\LTleft}{0pt}
\setlength{\LTright}{0pt}
\setlength{\tabcolsep}{6pt}
{\small
\begin{longtable}{@{}r l l l r r r r r r r r c l r r r r@{}}
\caption{MCVI scores for 185 economies (mean 2006--2025).}\label{tab:appendix}\\
\toprule
{Rank} & {Economy} & {Region} & {Classification} & {$N$} & {LSCI} & {Partners} & {HHI} & {MCVI} & {D1} & {D2} & {D3} & {Dom.} & {Cluster} & {MCVI} & {MCVI} & {$\Delta$\%} & {$\sigma$} \\
 & & & & & & & & & & & & & & {2006} & {2025} & & \\
\midrule
\endfirsthead

\toprule
{Rank} & {Economy} & {Region} & {Classification} & {$N$} & {LSCI} & {Partners} & {HHI} & {MCVI} & {D1} & {D2} & {D3} & {Dom.} & {Cluster} & {MCVI} & {MCVI} & {$\Delta$\%} & {$\sigma$} \\
 & & & & & & & & & & & & & & {2006} & {2025} & & \\
\midrule
\endhead

\endfoot

\bottomrule
\multicolumn{18}{l}{\scriptsize \textit{Notes:} Rank 1 = least vulnerable; Rank 185 = most vulnerable. Region follows the UN~M49 standard.}\\
\multicolumn{18}{l}{\scriptsize Classification: SIDS = Small Island Developing State; LDC = Least Developed Country; LLDC = Landlocked Developing Country.}\\
\multicolumn{18}{l}{\scriptsize $N$ = number of annual observations. LSCI = mean Liner Shipping Connectivity Index. Partners = mean number of bilateral partners. HHI = mean port Herfindahl--Hirschman Index.}\\
\multicolumn{18}{l}{\scriptsize D1 = Low connectivity (LSCI); D2 = Weak bilateral integration (LSBCI); D3 = Port concentration (HHI of PLSCI). MCVI = $(D_1 + D_2 + D_3)/3$.}\\
\multicolumn{18}{l}{\scriptsize Dom.\ = dimension with highest mean score. Cluster from $K$-means ($k=2$). $\Delta$\% = change between 2006 and 2025. $\sigma$ = temporal standard deviation. --- = not available.}\\
\multicolumn{18}{l}{\scriptsize Rankings for economies with $N < 5$ observations (e.g., Saint Helena, Saint Pierre and Miquelon, Falkland Islands, Cocos (Keeling) Islands) should be interpreted with caution.}\\
\multicolumn{18}{l}{\scriptsize Percent changes for low-MCVI economies are amplified by small denominators and should be read alongside $\sigma$.}\\
\endlastfoot

  1 & United States & Americas & --- & 20 & 486.7 & 159 & 0.048 & 0.015 & 0.022 & 0.017 & 0.006 & D1 & Well-connected & 0.021 & 0.011 & -45.9 & 0.007 \\
  2 & China & Asia & --- & 20 & 943.6 & 154 & 0.069 & 0.019 & 0.003 & 0.035 & 0.019 & D2 & Well-connected & 0.042 & 0.013 & -68.0 & 0.010 \\
  3 & United Kingdom & Europe & --- & 20 & 412.2 & 158 & 0.102 & 0.034 & 0.043 & 0.016 & 0.043 & D3 & Well-connected & 0.029 & 0.042 & +41.1 & 0.005 \\
  4 & Spain & Europe & --- & 20 & 381.1 & 156 & 0.111 & 0.044 & 0.054 & 0.027 & 0.050 & D1 & Well-connected & 0.049 & 0.035 & -28.6 & 0.006 \\
  5 & Japan & Asia & --- & 20 & 436.7 & 145 & 0.065 & 0.051 & 0.035 & 0.102 & 0.017 & D2 & Well-connected & 0.049 & 0.058 & +19.7 & 0.009 \\
  6 & Italy & Europe & --- & 20 & 294.6 & 154 & 0.091 & 0.053 & 0.074 & 0.049 & 0.035 & D1 & Well-connected & 0.058 & 0.050 & -14.4 & 0.004 \\
  7 & India & Asia & --- & 20 & 262.4 & 145 & 0.122 & 0.085 & 0.088 & 0.107 & 0.059 & D2 & Well-connected & 0.115 & 0.059 & -49.2 & 0.015 \\
  8 & France & Europe & --- & 20 & 283.7 & 155 & 0.223 & 0.088 & 0.077 & 0.035 & 0.151 & D3 & Well-connected & 0.089 & 0.094 & +5.1 & 0.007 \\
  9 & Malaysia & Asia & --- & 20 & 461.9 & 152 & 0.263 & 0.090 & 0.028 & 0.057 & 0.185 & D3 & Well-connected & 0.106 & 0.097 & -8.8 & 0.008 \\
  10 & Turkiye & Asia & --- & 20 & 219.4 & 139 & 0.099 & 0.101 & 0.112 & 0.148 & 0.043 & D2 & Well-connected & 0.151 & 0.056 & -62.9 & 0.033 \\
  11 & Republic of Korea & Asia & --- & 20 & 528.3 & 152 & 0.343 & 0.104 & 0.017 & 0.052 & 0.242 & D3 & Well-connected & 0.144 & 0.095 & -34.3 & 0.016 \\
  12 & Germany & Europe & --- & 20 & 390.9 & 154 & 0.347 & 0.115 & 0.048 & 0.049 & 0.247 & D3 & Well-connected & 0.121 & 0.125 & +2.8 & 0.008 \\
  13 & Mexico & Americas & --- & 20 & 165.2 & 141 & 0.166 & 0.130 & 0.152 & 0.134 & 0.103 & D1 & Well-connected & 0.157 & 0.114 & -27.6 & 0.018 \\
  14 & Brazil & Americas & --- & 20 & 138.3 & 135 & 0.078 & 0.131 & 0.185 & 0.181 & 0.026 & D1 & Well-connected & 0.112 & 0.111 & -1.1 & 0.013 \\
  15 & Portugal & Europe & --- & 20 & 158.0 & 143 & 0.177 & 0.132 & 0.160 & 0.122 & 0.113 & D1 & Well-connected & 0.175 & 0.114 & -34.9 & 0.018 \\
  16 & Egypt & Africa & --- & 20 & 229.6 & 142 & 0.261 & 0.135 & 0.103 & 0.118 & 0.182 & D3 & Well-connected & 0.163 & 0.100 & -38.8 & 0.017 \\
  17 & United Arab Emirates & Asia & --- & 20 & 289.4 & 146 & 0.343 & 0.139 & 0.075 & 0.097 & 0.244 & D3 & Well-connected & 0.139 & 0.132 & -5.0 & 0.007 \\
  18 & Canada & Americas & --- & 20 & 154.2 & 142 & 0.193 & 0.139 & 0.163 & 0.129 & 0.127 & D1 & Well-connected & 0.150 & 0.123 & -17.5 & 0.008 \\
  19 & Australia & Oceania & --- & 20 & 148.5 & 138 & 0.144 & 0.142 & 0.170 & 0.175 & 0.080 & D2 & Well-connected & 0.156 & 0.145 & -7.1 & 0.009 \\
  20 & Panama & Americas & --- & 20 & 178.6 & 142 & 0.252 & 0.149 & 0.142 & 0.130 & 0.175 & D3 & Well-connected & 0.218 & 0.110 & -49.5 & 0.033 \\
  21 & Russian Federation & Europe & --- & 20 & 180.4 & 134 & 0.174 & 0.157 & 0.142 & 0.219 & 0.110 & D2 & Well-connected & 0.231 & 0.185 & -19.9 & 0.050 \\
  22 & Colombia & Americas & --- & 20 & 166.9 & 141 & 0.260 & 0.159 & 0.151 & 0.144 & 0.182 & D3 & Well-connected & 0.207 & 0.137 & -33.8 & 0.032 \\
  23 & Saudi Arabia & Asia & --- & 20 & 212.8 & 143 & 0.357 & 0.160 & 0.114 & 0.113 & 0.252 & D3 & Well-connected & 0.210 & 0.128 & -39.0 & 0.030 \\
  24 & Indonesia & Asia & --- & 20 & 227.2 & 121 & 0.150 & 0.163 & 0.104 & 0.299 & 0.086 & D2 & Well-connected & 0.143 & 0.154 & +7.7 & 0.023 \\
  25 & Viet Nam & Asia & --- & 20 & 267.3 & 125 & 0.287 & 0.181 & 0.097 & 0.243 & 0.202 & D2 & Well-connected & 0.285 & 0.098 & -65.7 & 0.058 \\
  26 & South Africa & Africa & --- & 20 & 128.8 & 139 & 0.279 & 0.184 & 0.201 & 0.155 & 0.195 & D1 & Well-connected & 0.201 & 0.196 & -2.5 & 0.014 \\
  27 & Morocco & Africa & --- & 20 & 184.5 & 145 & 0.439 & 0.199 & 0.157 & 0.123 & 0.318 & D3 & Well-connected & 0.382 & 0.178 & -53.5 & 0.065 \\
  28 & Sweden & Europe & --- & 20 & 117.8 & 121 & 0.163 & 0.200 & 0.217 & 0.284 & 0.098 & D2 & Well-connected & 0.217 & 0.193 & -11.1 & 0.011 \\
  29 & New Zealand & Oceania & --- & 20 & 93.2 & 130 & 0.137 & 0.200 & 0.281 & 0.245 & 0.073 & D1 & Well-connected & 0.214 & 0.200 & -6.5 & 0.015 \\
  30 & Netherlands (Kingdom of the) & Europe & --- & 20 & 408.8 & 155 & 0.692 & 0.206 & 0.043 & 0.037 & 0.538 & D3 & Well-connected & 0.207 & 0.207 & -0.0 & 0.012 \\
  31 & Belgium & Europe & --- & 20 & 390.2 & 156 & 0.701 & 0.206 & 0.050 & 0.025 & 0.544 & D3 & Well-connected & 0.205 & 0.211 & +3.0 & 0.007 \\
  32 & Chile & Americas & --- & 20 & 94.0 & 121 & 0.116 & 0.209 & 0.279 & 0.294 & 0.054 & D2 & Well-connected & 0.238 & 0.202 & -15.1 & 0.019 \\
  33 & Thailand & Asia & --- & 20 & 215.4 & 128 & 0.444 & 0.213 & 0.113 & 0.213 & 0.313 & D3 & Well-connected & 0.199 & 0.191 & -4.3 & 0.016 \\
  34 & Dominican Republic & Americas & SIDS & 20 & 121.8 & 139 & 0.383 & 0.215 & 0.209 & 0.163 & 0.272 & D3 & Well-connected & 0.228 & 0.212 & -6.8 & 0.019 \\
  35 & Nigeria & Africa & --- & 20 & 90.4 & 131 & 0.226 & 0.229 & 0.291 & 0.244 & 0.153 & D1 & Well-connected & 0.265 & 0.185 & -29.9 & 0.022 \\
  36 & Oman & Asia & --- & 20 & 130.7 & 136 & 0.490 & 0.245 & 0.199 & 0.167 & 0.370 & D3 & Well-connected & 0.321 & 0.221 & -31.0 & 0.035 \\
  37 & Argentina & Americas & --- & 20 & 106.2 & 125 & 0.380 & 0.258 & 0.243 & 0.259 & 0.271 & D3 & Well-connected & 0.244 & 0.253 & +3.7 & 0.014 \\
  38 & Greece & Europe & --- & 20 & 147.8 & 135 & 0.557 & 0.262 & 0.179 & 0.174 & 0.433 & D3 & Well-connected & 0.229 & 0.252 & +10.0 & 0.015 \\
  39 & Pakistan & Asia & --- & 20 & 127.8 & 134 & 0.523 & 0.266 & 0.201 & 0.186 & 0.410 & D3 & Well-connected & 0.299 & 0.239 & -20.3 & 0.027 \\
  40 & Philippines & Asia & --- & 20 & 149.0 & 96 & 0.192 & 0.267 & 0.172 & 0.506 & 0.124 & D2 & Well-connected & 0.308 & 0.245 & -20.5 & 0.028 \\
  41 & Israel & Asia & --- & 20 & 108.0 & 130 & 0.498 & 0.273 & 0.240 & 0.223 & 0.358 & D3 & Well-connected & 0.275 & 0.330 & +19.8 & 0.022 \\
  42 & Peru & Americas & --- & 20 & 108.0 & 125 & 0.458 & 0.280 & 0.243 & 0.259 & 0.337 & D3 & Well-connected & 0.316 & 0.191 & -39.4 & 0.039 \\
  43 & Singapore & Asia & SIDS & 20 & 565.1 & 154 & 0.997 & 0.283 & 0.010 & 0.043 & 0.795 & D3 & Well-connected & 0.298 & 0.279 & -6.2 & 0.013 \\
  44 & Guatemala & Americas & --- & 20 & 93.3 & 121 & 0.368 & 0.288 & 0.282 & 0.319 & 0.264 & D2 & Well-connected & 0.270 & 0.297 & +9.8 & 0.030 \\
  45 & Denmark & Europe & --- & 20 & 92.0 & 114 & 0.327 & 0.291 & 0.286 & 0.353 & 0.233 & D2 & Well-connected & 0.342 & 0.269 & -21.4 & 0.022 \\
  46 & Angola & Africa & LDC & 20 & 68.5 & 124 & 0.312 & 0.300 & 0.374 & 0.304 & 0.222 & D1 & Well-connected & 0.322 & 0.306 & -5.0 & 0.030 \\
  47 & Finland & Europe & --- & 20 & 82.9 & 100 & 0.146 & 0.302 & 0.316 & 0.508 & 0.082 & D2 & Well-connected & 0.314 & 0.337 & +7.1 & 0.032 \\
  48 & Poland & Europe & --- & 20 & 97.5 & 115 & 0.399 & 0.302 & 0.288 & 0.334 & 0.284 & D2 & Well-connected & 0.442 & 0.247 & -44.1 & 0.074 \\
  49 & Venezuela (Bolivarian Rep. of) & Americas & --- & 20 & 72.2 & 114 & 0.190 & 0.303 & 0.401 & 0.385 & 0.123 & D1 & Well-connected & 0.222 & 0.424 & +91.4 & 0.110 \\
  50 & Malta & Europe & --- & 20 & 122.1 & 139 & 0.870 & 0.324 & 0.212 & 0.150 & 0.611 & D3 & Well-connected & 0.357 & 0.329 & -7.9 & 0.027 \\
  51 & Costa Rica & Americas & --- & 20 & 90.7 & 120 & 0.463 & 0.326 & 0.289 & 0.336 & 0.352 & D3 & Well-connected & 0.374 & 0.291 & -22.2 & 0.027 \\
  52 & Ecuador & Americas & --- & 20 & 95.4 & 124 & 0.583 & 0.334 & 0.277 & 0.274 & 0.451 & D3 & Well-connected & 0.409 & 0.235 & -42.5 & 0.065 \\
  53 & Jamaica & Americas & SIDS & 20 & 102.4 & 139 & 0.839 & 0.339 & 0.255 & 0.167 & 0.595 & D3 & Well-connected & 0.331 & 0.316 & -4.7 & 0.034 \\
  54 & Norway & Europe & --- & 20 & 58.8 & 79 & 0.047 & 0.347 & 0.423 & 0.611 & 0.006 & D2 & Well-connected & 0.409 & 0.381 & -6.8 & 0.032 \\
  55 & Sri Lanka & Asia & --- & 20 & 203.4 & 141 & 1.000 & 0.350 & 0.122 & 0.124 & 0.805 & D3 & Well-connected & 0.379 & 0.321 & -15.3 & 0.019 \\
  56 & Iran (Islamic Republic of) & Asia & --- & 20 & 83.8 & 108 & 0.440 & 0.350 & 0.329 & 0.394 & 0.327 & D2 & Well-connected & 0.358 & 0.395 & +10.4 & 0.055 \\
  57 & Ghana & Africa & --- & 20 & 89.4 & 130 & 0.646 & 0.351 & 0.296 & 0.242 & 0.514 & D3 & Well-connected & 0.372 & 0.310 & -16.8 & 0.017 \\
  58 & Trinidad and Tobago & Americas & SIDS & 20 & 71.9 & 125 & 0.514 & 0.357 & 0.361 & 0.321 & 0.391 & D3 & Well-connected & 0.368 & 0.419 & +14.0 & 0.043 \\
  59 & Algeria & Africa & --- & 20 & 64.2 & 93 & 0.203 & 0.359 & 0.392 & 0.549 & 0.135 & D2 & Well-connected & 0.337 & 0.350 & +4.0 & 0.029 \\
  60 & Cote d'Ivoire & Africa & --- & 20 & 83.1 & 124 & 0.624 & 0.372 & 0.317 & 0.298 & 0.502 & D3 & Well-connected & 0.374 & 0.325 & -13.0 & 0.021 \\
  61 & Bahamas & Americas & SIDS & 20 & 71.4 & 129 & 0.703 & 0.386 & 0.361 & 0.259 & 0.538 & D3 & Well-connected & 0.440 & 0.387 & -12.1 & 0.032 \\
  62 & Ukraine & Europe & --- & 19 & 62.2 & 103 & 0.449 & 0.388 & 0.424 & 0.415 & 0.327 & D1 & Well-connected & 0.435 & 0.639 & +46.9 & 0.127 \\
  63 & Libya & Africa & --- & 20 & 51.2 & 95 & 0.266 & 0.394 & 0.474 & 0.523 & 0.186 & D2 & Well-connected & 0.410 & 0.350 & -14.7 & 0.041 \\
  64 & Mozambique & Africa & LDC & 20 & 45.1 & 106 & 0.307 & 0.396 & 0.518 & 0.454 & 0.217 & D1 & Well-connected & 0.474 & 0.345 & -27.1 & 0.065 \\
  65 & Papua New Guinea & Oceania & SIDS & 20 & 41.9 & 89 & 0.170 & 0.407 & 0.539 & 0.578 & 0.105 & D2 & Well-connected & 0.408 & 0.388 & -5.0 & 0.029 \\
  66 & Lebanon & Asia & --- & 20 & 87.9 & 119 & 0.848 & 0.427 & 0.299 & 0.298 & 0.682 & D3 & Well-connected & 0.495 & 0.370 & -25.2 & 0.076 \\
  67 & Ireland & Europe & --- & 20 & 58.7 & 98 & 0.474 & 0.432 & 0.428 & 0.523 & 0.344 & D2 & Well-connected & 0.403 & 0.403 & +0.2 & 0.036 \\
  68 & Uruguay & Americas & --- & 20 & 85.0 & 120 & 0.957 & 0.437 & 0.310 & 0.307 & 0.695 & D3 & Well-connected & 0.511 & 0.390 & -23.7 & 0.053 \\
  69 & Qatar & Asia & --- & 20 & 62.1 & 98 & 0.594 & 0.441 & 0.443 & 0.454 & 0.427 & D2 & Well-connected & 0.438 & 0.392 & -10.4 & 0.060 \\
  70 & United Republic of Tanzania & Africa & --- & 20 & 54.0 & 104 & 0.563 & 0.453 & 0.453 & 0.466 & 0.440 & D2 & Well-connected & 0.488 & 0.460 & -5.7 & 0.039 \\
  71 & Cameroon & Africa & --- & 20 & 64.4 & 123 & 0.816 & 0.454 & 0.390 & 0.319 & 0.653 & D3 & Well-connected & 0.503 & 0.366 & -27.4 & 0.059 \\
  72 & Honduras & Americas & --- & 20 & 61.3 & 98 & 0.555 & 0.457 & 0.407 & 0.518 & 0.445 & D2 & Well-connected & 0.423 & 0.475 & +12.3 & 0.034 \\
  73 & Fiji & Oceania & SIDS & 20 & 42.7 & 106 & 0.506 & 0.458 & 0.528 & 0.468 & 0.376 & D1 & Well-connected & 0.527 & 0.478 & -9.3 & 0.046 \\
  74 & Croatia & Europe & --- & 20 & 52.7 & 102 & 0.612 & 0.464 & 0.463 & 0.451 & 0.479 & D3 & Well-connected & 0.572 & 0.481 & -15.9 & 0.046 \\
  75 & Gabon & Africa & --- & 20 & 38.3 & 106 & 0.508 & 0.470 & 0.565 & 0.459 & 0.386 & D1 & Well-connected & 0.456 & 0.502 & +10.3 & 0.030 \\
  76 & Tunisia & Africa & --- & 20 & 41.9 & 58 & 0.278 & 0.478 & 0.547 & 0.696 & 0.192 & D2 & Well-connected & 0.449 & 0.570 & +27.0 & 0.064 \\
  77 & Madagascar & Africa & LDC & 20 & 32.4 & 72 & 0.255 & 0.484 & 0.628 & 0.649 & 0.175 & D2 & Well-connected & 0.421 & 0.511 & +21.3 & 0.050 \\
  78 & Togo & Africa & LDC & 20 & 73.4 & 124 & 1.000 & 0.486 & 0.366 & 0.287 & 0.805 & D3 & Well-connected & 0.549 & 0.398 & -27.4 & 0.050 \\
  79 & Mauritius & Africa & SIDS & 20 & 66.0 & 115 & 0.962 & 0.487 & 0.384 & 0.355 & 0.724 & D3 & Well-connected & 0.529 & 0.431 & -18.5 & 0.043 \\
  80 & Djibouti & Africa & LDC & 20 & 70.4 & 123 & 1.000 & 0.488 & 0.369 & 0.291 & 0.805 & D3 & Well-connected & 0.543 & 0.498 & -8.3 & 0.038 \\
  81 & Guadeloupe & Americas & --- & 20 & 40.7 & 113 & 0.639 & 0.493 & 0.545 & 0.419 & 0.514 & D1 & Well-connected & 0.465 & 0.504 & +8.5 & 0.048 \\
  82 & Congo & Africa & --- & 20 & 66.6 & 119 & 1.000 & 0.510 & 0.388 & 0.337 & 0.805 & D3 & Vulnerable & 0.549 & 0.434 & -21.0 & 0.036 \\
  83 & Jordan & Asia & --- & 20 & 66.5 & 113 & 1.000 & 0.514 & 0.379 & 0.358 & 0.805 & D3 & Vulnerable & 0.504 & 0.500 & -0.8 & 0.015 \\
  84 & Kuwait & Asia & --- & 20 & 40.3 & 85 & 0.519 & 0.514 & 0.554 & 0.593 & 0.395 & D2 & Vulnerable & 0.428 & 0.513 & +20.1 & 0.057 \\
  85 & Namibia & Africa & --- & 20 & 44.4 & 107 & 0.808 & 0.515 & 0.524 & 0.425 & 0.597 & D3 & Vulnerable & 0.567 & 0.407 & -28.2 & 0.066 \\
  86 & Benin & Africa & LDC & 20 & 57.0 & 120 & 1.000 & 0.525 & 0.434 & 0.338 & 0.805 & D3 & Vulnerable & 0.568 & 0.499 & -12.1 & 0.034 \\
  87 & New Caledonia & Oceania & SIDS & 20 & 44.9 & 121 & 0.897 & 0.526 & 0.512 & 0.359 & 0.707 & D3 & Vulnerable & 0.573 & 0.468 & -18.3 & 0.041 \\
  88 & Equatorial Guinea & Africa & --- & 20 & 30.8 & 83 & 0.502 & 0.529 & 0.647 & 0.563 & 0.376 & D1 & Vulnerable & 0.588 & 0.506 & -13.9 & 0.052 \\
  89 & Slovenia & Europe & --- & 20 & 64.9 & 105 & 1.000 & 0.539 & 0.392 & 0.419 & 0.805 & D3 & Vulnerable & 0.573 & 0.514 & -10.2 & 0.035 \\
  90 & Syrian Arab Republic & Asia & --- & 20 & 40.1 & 93 & 0.644 & 0.544 & 0.565 & 0.556 & 0.510 & D1 & Vulnerable & 0.463 & 0.573 & +23.7 & 0.056 \\
  91 & Romania & Europe & --- & 20 & 63.8 & 105 & 1.000 & 0.544 & 0.395 & 0.433 & 0.805 & D3 & Vulnerable & 0.551 & 0.591 & +7.1 & 0.026 \\
  92 & Senegal & Africa & LDC & 20 & 56.9 & 113 & 1.000 & 0.545 & 0.434 & 0.397 & 0.805 & D3 & Vulnerable & 0.530 & 0.535 & +0.9 & 0.023 \\
  93 & Cuba & Americas & SIDS & 20 & 30.6 & 73 & 0.492 & 0.545 & 0.645 & 0.620 & 0.371 & D1 & Vulnerable & 0.589 & 0.600 & +1.8 & 0.034 \\
  94 & Estonia & Europe & --- & 20 & 35.4 & 65 & 0.481 & 0.550 & 0.595 & 0.685 & 0.369 & D2 & Vulnerable & 0.633 & 0.606 & -4.2 & 0.057 \\
  95 & Kenya & Africa & --- & 20 & 60.5 & 106 & 1.000 & 0.555 & 0.413 & 0.447 & 0.805 & D3 & Vulnerable & 0.548 & 0.526 & -4.0 & 0.016 \\
  96 & Cyprus & Europe & --- & 20 & 55.7 & 102 & 0.939 & 0.556 & 0.442 & 0.481 & 0.745 & D3 & Vulnerable & 0.584 & 0.610 & +4.4 & 0.054 \\
  97 & Yemen & Asia & LDC & 20 & 37.0 & 64 & 0.503 & 0.558 & 0.610 & 0.671 & 0.392 & D2 & Vulnerable & 0.371 & 0.519 & +39.8 & 0.135 \\
  98 & Bangladesh & Asia & LDC & 20 & 56.6 & 70 & 0.781 & 0.561 & 0.445 & 0.670 & 0.570 & D2 & Vulnerable & 0.535 & 0.502 & -6.1 & 0.054 \\
  99 & French Polynesia & Oceania & SIDS & 20 & 39.0 & 111 & 0.938 & 0.562 & 0.558 & 0.429 & 0.699 & D3 & Vulnerable & 0.586 & 0.527 & -10.1 & 0.039 \\
  100 & Iceland & Europe & --- & 20 & 20.5 & 29 & 0.214 & 0.569 & 0.783 & 0.783 & 0.142 & D1 & Vulnerable & 0.593 & 0.487 & -17.9 & 0.041 \\
  101 & Reunion & Africa & --- & 20 & 49.6 & 107 & 1.000 & 0.571 & 0.483 & 0.425 & 0.805 & D3 & Vulnerable & 0.566 & 0.528 & -6.6 & 0.039 \\
  102 & Puerto Rico & Americas & SIDS & 20 & 52.2 & 105 & 0.989 & 0.575 & 0.466 & 0.474 & 0.784 & D3 & Vulnerable & 0.444 & 0.638 & +43.5 & 0.060 \\
  103 & Haiti & Americas & SIDS, LDC & 20 & 33.9 & 66 & 0.622 & 0.591 & 0.616 & 0.687 & 0.469 & D2 & Vulnerable & 0.819 & 0.532 & -35.0 & 0.107 \\
  104 & Vanuatu & Oceania & SIDS, LDC & 20 & 23.7 & 66 & 0.519 & 0.605 & 0.733 & 0.679 & 0.404 & D1 & Vulnerable & 0.557 & 0.584 & +4.9 & 0.050 \\
  105 & Iraq & Asia & --- & 20 & 43.4 & 56 & 0.743 & 0.608 & 0.584 & 0.676 & 0.565 & D2 & Vulnerable & 0.889 & 0.419 & -52.9 & 0.156 \\
  106 & Sint Maarten (Dutch part) & Americas & --- & 15 & 37.0 & 107 & 1.000 & 0.617 & 0.578 & 0.469 & 0.805 & D3 & Vulnerable & --- & 0.658 & --- & 0.030 \\
  107 & Bulgaria & Europe & --- & 20 & 27.1 & 57 & 0.555 & 0.618 & 0.695 & 0.718 & 0.440 & D2 & Vulnerable & 0.507 & 0.562 & +10.8 & 0.067 \\
  108 & Martinique & Americas & --- & 20 & 35.9 & 106 & 1.000 & 0.618 & 0.590 & 0.459 & 0.805 & D3 & Vulnerable & 0.588 & 0.602 & +2.5 & 0.038 \\
  109 & Lithuania & Europe & --- & 20 & 48.0 & 88 & 1.000 & 0.623 & 0.499 & 0.564 & 0.805 & D3 & Vulnerable & 0.743 & 0.510 & -31.4 & 0.063 \\
  110 & Solomon Islands & Oceania & SIDS, LDC & 20 & 29.6 & 80 & 0.766 & 0.628 & 0.661 & 0.619 & 0.603 & D1 & Vulnerable & 0.752 & 0.607 & -19.2 & 0.066 \\
  111 & Saint Lucia & Americas & SIDS & 20 & 26.7 & 68 & 0.693 & 0.630 & 0.694 & 0.647 & 0.548 & D1 & Vulnerable & 0.592 & 0.717 & +21.1 & 0.089 \\
  112 & Somalia & Africa & LDC & 19 & 24.3 & 38 & 0.457 & 0.631 & 0.734 & 0.813 & 0.345 & D2 & Vulnerable & 0.715 & 0.443 & -38.1 & 0.107 \\
  113 & Suriname & Americas & SIDS & 20 & 37.9 & 98 & 1.000 & 0.631 & 0.568 & 0.520 & 0.805 & D3 & Vulnerable & 0.648 & 0.612 & -5.6 & 0.021 \\
  114 & Guyana & Americas & SIDS & 20 & 38.3 & 97 & 1.000 & 0.632 & 0.564 & 0.526 & 0.805 & D3 & Vulnerable & 0.648 & 0.644 & -0.5 & 0.029 \\
  115 & Mauritania & Africa & LDC & 20 & 25.5 & 53 & 0.577 & 0.633 & 0.708 & 0.722 & 0.468 & D2 & Vulnerable & 0.657 & 0.627 & -4.5 & 0.059 \\
  116 & Dem. Rep. of the Congo & Africa & --- & 20 & 23.2 & 40 & 0.524 & 0.634 & 0.744 & 0.746 & 0.413 & D2 & Vulnerable & 0.662 & 0.689 & +4.2 & 0.073 \\
  117 & Saint Vincent and the Grenadines & Americas & SIDS & 20 & 25.9 & 65 & 0.681 & 0.637 & 0.706 & 0.659 & 0.544 & D1 & Vulnerable & 0.730 & 0.649 & -11.1 & 0.071 \\
  118 & Bahrain & Asia & SIDS & 20 & 43.1 & 85 & 1.000 & 0.637 & 0.535 & 0.570 & 0.805 & D3 & Vulnerable & 0.659 & 0.722 & +9.6 & 0.059 \\
  119 & Nicaragua & Americas & --- & 20 & 25.0 & 60 & 0.667 & 0.643 & 0.714 & 0.706 & 0.509 & D1 & Vulnerable & 0.808 & 0.653 & -19.2 & 0.072 \\
  120 & Cabo Verde & Africa & SIDS & 20 & 15.9 & 22 & 0.387 & 0.651 & 0.844 & 0.829 & 0.280 & D1 & Vulnerable & 0.675 & 0.668 & -1.1 & 0.034 \\
  121 & Barbados & Americas & SIDS & 20 & 36.0 & 89 & 1.000 & 0.651 & 0.586 & 0.562 & 0.805 & D3 & Vulnerable & 0.639 & 0.665 & +4.0 & 0.028 \\
  122 & Latvia & Europe & --- & 20 & 33.9 & 64 & 0.862 & 0.653 & 0.610 & 0.682 & 0.668 & D2 & Vulnerable & 0.641 & 0.552 & -13.9 & 0.059 \\
  123 & Aruba & Americas & SIDS & 20 & 27.7 & 72 & 0.836 & 0.662 & 0.683 & 0.635 & 0.669 & D1 & Vulnerable & 0.675 & 0.712 & +5.6 & 0.049 \\
  124 & El Salvador & Americas & --- & 20 & 27.1 & 68 & 0.773 & 0.664 & 0.687 & 0.682 & 0.625 & D1 & Vulnerable & 0.712 & 0.776 & +9.0 & 0.074 \\
  125 & United States Virgin Islands & Americas & SIDS & 20 & 21.2 & 42 & 0.573 & 0.665 & 0.773 & 0.774 & 0.448 & D2 & Vulnerable & 0.577 & 0.696 & +20.6 & 0.058 \\
  126 & Guinea & Africa & LDC & 20 & 32.2 & 91 & 1.000 & 0.666 & 0.630 & 0.562 & 0.805 & D3 & Vulnerable & 0.614 & 0.668 & +8.7 & 0.055 \\
  127 & Belize & Americas & SIDS & 20 & 21.5 & 48 & 0.647 & 0.671 & 0.765 & 0.732 & 0.514 & D1 & Vulnerable & 0.844 & 0.662 & -21.5 & 0.110 \\
  128 & Micronesia (Federated States of) & Oceania & --- & 20 & 10.7 & 15 & 0.329 & 0.672 & 0.907 & 0.875 & 0.236 & D1 & Vulnerable & 0.686 & 0.626 & -8.7 & 0.038 \\
  129 & Curacao & Americas & --- & 15 & 31.2 & 84 & 1.000 & 0.680 & 0.638 & 0.598 & 0.805 & D3 & Vulnerable & --- & 0.729 & --- & 0.034 \\
  130 & Myanmar & Asia & LDC & 20 & 34.0 & 46 & 0.862 & 0.697 & 0.626 & 0.767 & 0.697 & D2 & Vulnerable & 0.852 & 0.586 & -31.2 & 0.113 \\
  131 & Tonga & Oceania & SIDS & 20 & 24.5 & 55 & 0.827 & 0.699 & 0.724 & 0.716 & 0.658 & D1 & Vulnerable & 0.740 & 0.624 & -15.7 & 0.073 \\
  132 & Guam & Oceania & SIDS & 20 & 23.9 & 69 & 0.885 & 0.702 & 0.732 & 0.661 & 0.714 & D1 & Vulnerable & 0.763 & 0.635 & -16.8 & 0.053 \\
  133 & Samoa & Oceania & SIDS & 20 & 28.7 & 69 & 1.000 & 0.703 & 0.668 & 0.635 & 0.805 & D3 & Vulnerable & 0.655 & 0.691 & +5.5 & 0.040 \\
  134 & American Samoa & Oceania & SIDS & 20 & 26.6 & 67 & 1.000 & 0.713 & 0.695 & 0.640 & 0.805 & D3 & Vulnerable & 0.676 & 0.706 & +4.5 & 0.047 \\
  135 & Comoros & Africa & SIDS, LDC & 20 & 16.0 & 26 & 0.613 & 0.716 & 0.843 & 0.829 & 0.474 & D1 & Vulnerable & 0.854 & 0.709 & -17.1 & 0.086 \\
  136 & Marshall Islands & Oceania & SIDS & 20 & 18.3 & 42 & 0.698 & 0.718 & 0.812 & 0.773 & 0.569 & D1 & Vulnerable & 0.853 & 0.667 & -21.9 & 0.077 \\
  137 & Georgia & Asia & --- & 20 & 22.0 & 21 & 0.675 & 0.718 & 0.762 & 0.853 & 0.539 & D2 & Vulnerable & 0.807 & 0.622 & -22.8 & 0.061 \\
  138 & Seychelles & Africa & SIDS & 20 & 24.5 & 65 & 0.911 & 0.721 & 0.723 & 0.707 & 0.734 & D3 & Vulnerable & 0.671 & 0.722 & +7.7 & 0.040 \\
  139 & Grenada & Americas & SIDS & 20 & 24.9 & 64 & 1.000 & 0.730 & 0.719 & 0.667 & 0.805 & D3 & Vulnerable & 0.743 & 0.749 & +0.9 & 0.035 \\
  140 & Sudan & Africa & LDC & 14 & 28.6 & 54 & 1.000 & 0.740 & 0.673 & 0.742 & 0.805 & D3 & Vulnerable & --- & 0.781 & --- & 0.070 \\
  141 & Cambodia & Asia & LDC & 20 & 28.7 & 48 & 1.000 & 0.748 & 0.684 & 0.754 & 0.805 & D3 & Vulnerable & 0.838 & 0.654 & -22.0 & 0.067 \\
  142 & Antigua and Barbuda & Americas & SIDS & 20 & 22.8 & 51 & 1.000 & 0.754 & 0.746 & 0.711 & 0.805 & D3 & Vulnerable & 0.737 & 0.725 & -1.5 & 0.038 \\
  143 & Saint Kitts and Nevis & Americas & SIDS & 20 & 18.3 & 36 & 0.823 & 0.755 & 0.809 & 0.788 & 0.666 & D1 & Vulnerable & 0.845 & 0.878 & +3.9 & 0.092 \\
  144 & Dominica & Americas & SIDS & 20 & 21.5 & 47 & 1.000 & 0.765 & 0.767 & 0.724 & 0.805 & D3 & Vulnerable & 0.741 & 0.741 & +0.0 & 0.027 \\
  145 & British Virgin Islands & Americas & SIDS & 20 & 13.2 & 20 & 0.755 & 0.769 & 0.875 & 0.843 & 0.588 & D1 & Vulnerable & 0.839 & 0.878 & +4.6 & 0.102 \\
  146 & Saint Pierre and Miquelon & Americas & --- & 2 & 3.8 & 1 & 0.500 & 0.772 & 0.988 & 0.983 & 0.345 & D1 & Vulnerable & --- & 0.772 & --- & 0.000 \\
  147 & Faroe Islands & Europe & --- & 20 & 15.3 & 16 & 0.762 & 0.772 & 0.854 & 0.846 & 0.618 & D1 & Vulnerable & 0.851 & 0.687 & -19.3 & 0.067 \\
  148 & Mayotte & Africa & --- & 20 & 20.6 & 48 & 1.000 & 0.779 & 0.779 & 0.753 & 0.805 & D3 & Vulnerable & 0.738 & 0.831 & +12.5 & 0.056 \\
  149 & Sierra Leone & Africa & LDC & 20 & 20.2 & 45 & 1.000 & 0.781 & 0.785 & 0.753 & 0.805 & D3 & Vulnerable & 0.856 & 0.775 & -9.5 & 0.053 \\
  150 & Gambia & Africa & LDC & 20 & 19.7 & 40 & 1.000 & 0.786 & 0.792 & 0.762 & 0.805 & D3 & Vulnerable & 0.866 & 0.852 & -1.7 & 0.073 \\
  151 & Liberia & Africa & LDC & 20 & 19.7 & 37 & 1.000 & 0.790 & 0.792 & 0.775 & 0.805 & D3 & Vulnerable & 0.856 & 0.782 & -8.7 & 0.044 \\
  152 & French Guiana & Americas & --- & 20 & 17.0 & 43 & 1.000 & 0.792 & 0.832 & 0.739 & 0.805 & D1 & Vulnerable & 0.736 & 0.765 & +3.9 & 0.020 \\
  153 & Brunei Darussalam & Asia & --- & 20 & 22.8 & 25 & 1.000 & 0.794 & 0.748 & 0.830 & 0.805 & D2 & Vulnerable & 0.813 & 0.798 & -1.9 & 0.027 \\
  154 & Maldives & Asia & SIDS & 20 & 18.3 & 34 & 0.978 & 0.803 & 0.814 & 0.807 & 0.788 & D1 & Vulnerable & 0.753 & 0.795 & +5.5 & 0.040 \\
  155 & Cook Islands & Oceania & SIDS & 19 & 7.8 & 4 & 0.670 & 0.806 & 0.938 & 0.946 & 0.533 & D2 & Vulnerable & 0.924 & 0.756 & -18.2 & 0.075 \\
  156 & Greenland & Europe & --- & 14 & 7.6 & 2 & 0.721 & 0.826 & 0.941 & 0.972 & 0.566 & D2 & Vulnerable & 0.924 & 0.784 & -15.1 & 0.073 \\
  157 & Sao Tome and Principe & Africa & SIDS, LDC & 20 & 13.6 & 23 & 1.000 & 0.835 & 0.874 & 0.825 & 0.805 & D1 & Vulnerable & 0.834 & 0.859 & +3.0 & 0.034 \\
  158 & Kiribati & Oceania & SIDS, LDC & 17 & 13.5 & 29 & 1.000 & 0.836 & 0.873 & 0.831 & 0.805 & D1 & Vulnerable & --- & 0.808 & --- & 0.036 \\
  159 & Northern Mariana Islands & Oceania & --- & 20 & 14.4 & 21 & 1.000 & 0.838 & 0.863 & 0.847 & 0.805 & D1 & Vulnerable & 0.869 & 0.837 & -3.6 & 0.026 \\
  160 & Anguilla & Americas & SIDS & 11 & 13.0 & 23 & 1.000 & 0.839 & 0.879 & 0.831 & 0.805 & D1 & Vulnerable & 0.878 & --- & --- & 0.027 \\
  161 & Montserrat & Americas & SIDS & 8 & 12.7 & 23 & 1.000 & 0.841 & 0.885 & 0.832 & 0.805 & D1 & Vulnerable & --- & --- & --- & 0.015 \\
  162 & Guinea-Bissau & Africa & SIDS, LDC & 20 & 12.9 & 12 & 1.000 & 0.851 & 0.881 & 0.867 & 0.805 & D1 & Vulnerable & 0.865 & 0.838 & -3.1 & 0.023 \\
  163 & Albania & Europe & --- & 19 & 13.8 & 12 & 1.000 & 0.853 & 0.871 & 0.882 & 0.805 & D2 & Vulnerable & --- & 0.859 & --- & 0.025 \\
  164 & Montenegro & Europe & --- & 18 & 12.6 & 12 & 1.000 & 0.861 & 0.884 & 0.895 & 0.805 & D2 & Vulnerable & --- & 0.798 & --- & 0.027 \\
  165 & Gibraltar & Europe & --- & 18 & 8.1 & 8 & 1.000 & 0.871 & 0.935 & 0.874 & 0.805 & D1 & Vulnerable & 0.876 & 0.881 & +0.6 & 0.020 \\
  166 & Palau & Oceania & SIDS & 19 & 9.0 & 7 & 1.000 & 0.875 & 0.923 & 0.896 & 0.805 & D1 & Vulnerable & 0.869 & 0.895 & +3.1 & 0.013 \\
  167 & Timor-Leste & Asia & SIDS, LDC & 20 & 10.7 & 5 & 1.000 & 0.878 & 0.906 & 0.924 & 0.805 & D2 & Vulnerable & 0.888 & 0.870 & -2.0 & 0.029 \\
  168 & Cayman Islands & Americas & SIDS & 20 & 8.5 & 3 & 1.000 & 0.888 & 0.928 & 0.932 & 0.805 & D2 & Vulnerable & 0.883 & 0.891 & +1.0 & 0.010 \\
  169 & Turks and Caicos Islands & Americas & SIDS & 10 & 4.6 & 2 & 0.911 & 0.890 & 0.980 & 0.955 & 0.735 & D1 & Vulnerable & 0.786 & 0.917 & +16.5 & 0.054 \\
  170 & Eritrea & Africa & LDC & 16 & 8.0 & 5 & 1.000 & 0.892 & 0.935 & 0.937 & 0.805 & D2 & Vulnerable & 0.830 & 0.904 & +8.9 & 0.026 \\
  171 & Bermuda & Americas & SIDS & 20 & 7.8 & 1 & 1.000 & 0.901 & 0.936 & 0.963 & 0.805 & D2 & Vulnerable & 0.900 & 0.902 & +0.3 & 0.001 \\
  172 & Wallis and Futuna Islands & Oceania & SIDS & 18 & 6.5 & 5 & 1.000 & 0.902 & 0.951 & 0.949 & 0.805 & D1 & Vulnerable & 0.881 & 0.913 & +3.6 & 0.014 \\
  173 & Tuvalu & Oceania & SIDS, LDC & 16 & 6.2 & 4 & 1.000 & 0.906 & 0.956 & 0.956 & 0.805 & D1 & Vulnerable & --- & 0.924 & --- & 0.017 \\
  174 & Guernsey & Europe & --- & 11 & 5.5 & 2 & 1.000 & 0.906 & 0.964 & 0.950 & 0.805 & D1 & Vulnerable & 0.909 & --- & --- & 0.007 \\
  174 & Jersey & Europe & --- & 11 & 5.5 & 2 & 1.000 & 0.906 & 0.964 & 0.950 & 0.805 & D1 & Vulnerable & 0.909 & --- & --- & 0.007 \\
  176 & Christmas Island & Oceania & --- & 10 & 4.8 & 3 & 1.000 & 0.907 & 0.974 & 0.941 & 0.805 & D1 & Vulnerable & --- & 0.902 & --- & 0.005 \\
  177 & Niue & Oceania & SIDS & 8 & 5.3 & 4 & 1.000 & 0.909 & 0.966 & 0.955 & 0.805 & D1 & Vulnerable & --- & 0.904 & --- & 0.006 \\
  178 & Paraguay & Americas & LLDC & 18 & 6.2 & 2 & 1.000 & 0.912 & 0.956 & 0.976 & 0.805 & D2 & Vulnerable & --- & 0.901 & --- & 0.009 \\
  179 & Nauru & Oceania & SIDS & 9 & 4.9 & 3 & 1.000 & 0.915 & 0.974 & 0.966 & 0.805 & D1 & Vulnerable & --- & --- & --- & 0.008 \\
  180 & Cocos (Keeling) Islands & Oceania & --- & 2 & 3.5 & 2 & 1.000 & 0.919 & 0.991 & 0.961 & 0.805 & D1 & Vulnerable & 0.919 & --- & --- & 0.000 \\
  181 & Falkland Islands (Malvinas) & Americas & --- & 2 & 4.3 & 2 & 1.000 & 0.920 & 0.982 & 0.972 & 0.805 & D1 & Vulnerable & --- & --- & --- & 0.000 \\
  182 & Norfolk Island & Oceania & --- & 20 & 3.7 & 2 & 1.000 & 0.923 & 0.985 & 0.979 & 0.805 & D1 & Vulnerable & 0.928 & 0.931 & +0.3 & 0.015 \\
  183 & Saint Helena & Africa & --- & 1 & 3.4 & 2 & 1.000 & 0.923 & 0.991 & 0.973 & 0.805 & D1 & Vulnerable & --- & 0.923 & --- & --- \\
  184 & Dem. People's Rep. of Korea & Asia & --- & 7 & 3.2 & 1 & 1.000 & 0.924 & 0.993 & 0.975 & 0.805 & D1 & Vulnerable & 0.926 & --- & --- & 0.001 \\
  185 & Republic of Moldova & Europe & LLDC & 10 & 3.4 & 1 & 1.000 & 0.929 & 0.991 & 0.992 & 0.805 & D2 & Vulnerable & --- & --- & --- & 0.001 \\

\end{longtable}
}
\end{landscape}
\restoregeometry
\pagestyle{fancy}

\bibliographystyle{unsrtnat}
\bibliography{Bib/mcvi_references}

\end{document}